\documentclass[acmtog]{acmart}
\acmSubmissionID{180}

\def\BibTeX{{\rm B\kern-.05em{\sc i\kern-.025em b}\kern-.08emT\kern-.1667em\lower.7ex\hbox{E}\kern-.125emX}}

\setcitestyle{square}

\usepackage{wrapfig}
\usepackage{nicefrac}
\usepackage{mathtools}
\usepackage{graphicx}
\usepackage{nicefrac}
\usepackage{booktabs} 
\usepackage{colortbl}
\usepackage{tabularx}
\usepackage{float}
\usepackage[ruled,vlined]{algorithm2e}
\usepackage{amsmath}
\usepackage{amsfonts}
\usepackage{amssymb}
\usepackage{amsbsy}

\definecolor{myBlue}{RGB}{144,210,236}
\definecolor{myGray}{RGB}{200,200,200}
\definecolor{iglGreen}{RGB}{153,203,67}
\definecolor{gaBlue}{RGB}{221, 65, 50}

\newcommand{\update}[1] {#1}

\newcommand{\refequ}[1] {Eq.~\ref{equ:#1}}
\newcommand{\reffig}[1] {Fig.~\ref{fig:#1}}

\newcommand{\refsec}[1] {Section~\ref{sec:#1}}
\newcommand{\refapp}[1] {Appendix~\ref{app:#1}}
\newcommand{\refalg}[1] {Algorithm~\ref{alg:#1}}

\newcommand{\mass}{\mathsf{M}} 
\newcommand{\massc}{\mathsf{M}_c} 
\newcommand{\lap}{\mathsf{L}} 
\newcommand{\nadam}{\textsc{nadam}\xspace} 
\newcommand{\fmap}{\mathsf{C}} 
\newcommand{\refshape}{\mathcal{R}} 
\newcommand{\defshape}{\mathcal{D}} 
\newcommand{\shapediff}{\mathsf{A}} 
\newcommand{\modshapediff}{\tilde{\mathsf{A}}} 

\newcommand{\eVec}{{\Phi}} 
\newcommand{\R}{\mathbb{R}}
\newcommand{\id}{\textsf{id}} 

\def\vb{{\mathbf{b}}}

\def\vg{{\mathbf{g}}}

\usepackage[mathletters]{ucs}
\usepackage[utf8x]{inputenc}
\newcommand{\coarse}[1]{\widetilde{#1}}
\newcommand{\mat}[1]{\mathsf{#1}}
\renewcommand{\L}{\mat{L}}
\newcommand{\Lc}{\coarse{\mat{L}}}
\renewcommand{\Mc}{\coarse{\mat{M}}}
\renewcommand{\P}{\mat{P}}

\newcommand{\K}{\mat{K}}

\renewcommand{\C}{\mat{C}}
\newcommand{\M}{\mat{M}}
\newcommand{\I}{\mat{I}}
\newcommand{\D}{\mat{D}}
\newcommand{\X}{\mat{X}}
\newcommand{\Phic}{\coarse{Φ}}
\renewcommand{\G}{\mat{G}}
\newcommand{\A}{\mat{A}}

\newcommand{\SL}{{\mat{S}_{\L}}}
\newcommand{\SLc}{{\mat{S}_{\Lc}}}
\newcommand{\SG}{{\mat{S}_{\G}}}
\newcommand{\Ac}{{\coarse{\mat{A}}}}

\newcommand{\src}{\mathcal{N}} 
\newcommand{\tar}{\mathcal{M}} 
\newcommand{\srcc}{{\coarse{\mathcal{N}}}}
\newcommand{\tarc}{\coarse{\mathcal{M}}} 
\newcommand{\T}{\mat{T}} 

\usepackage[verbose]{newunicodechar}

\newcommand{\Z}{\mat{Z}}

\newcommand{\f}{\mathbf{f}}

\newcommand{\vectorize}{\mathop{\text{vec}}}

\begin{document}
\title{Spectral Coarsening of Geometric Operators}
\author{Hsueh-Ti Derek Liu}
\affiliation{
	\institution{University of Toronto}
	\streetaddress{40 St. George Street}
 	\city{Toronto}
 	\state{ON}
 	\postcode{M5S 2E4}
 	\country{Canada}}
\author{Alec Jacobson}
\affiliation{
	\institution{University of Toronto}
	\streetaddress{40 St. George Street}
 	\city{Toronto}
 	\state{ON}
 	\postcode{M5S 2E4}
 	\country{Canada}}
\author{Maks Ovsjanikov}
\affiliation{
	\institution{\'Ecole Polytechnique}
	\streetaddress{1 Rue Honor\'e d'Estienne d'Orves}
 	\city{Palaiseau}
 	\postcode{91120}
 	\country{France}}

\begin{abstract}
  We introduce a novel approach to measure the behavior of a geometric operator
  before and after coarsening.
  By comparing eigenvectors of the input operator and its coarsened counterpart,
  we can quantitatively and visually analyze how well the spectral properties of
  the operator are maintained.
  Using this measure, we show that standard mesh simplification and algebraic
  coarsening techniques fail to maintain spectral properties.
  In response, 
  we introduce a novel approach for \emph{spectral coarsening}.
  We show that it is possible to significantly reduce the sampling density of an
  operator derived from a 3D shape without affecting the low-frequency
  eigenvectors.
  By marrying techniques developed within the algebraic multigrid and the
  functional maps literatures, we successfully coarsen a variety of isotropic
  and anisotropic operators while maintaining sparsity and positive
  semi-definiteness.
  We demonstrate the utility of this approach for applications including operator-sensitive sampling, shape matching, and graph pooling for convolutional neural networks.
\end{abstract}

\ccsdesc[500]{Computing methodologies~Shape analysis}
\ccsdesc[300]{Mathematics of computing~Computations on matrices}

\setcopyright{acmlicensed}
\acmJournal{TOG}
\acmYear{2019}\acmVolume{38}\acmNumber{4}\acmArticle{1}\acmMonth{7} \acmDOI{10.1145/3306346.3322953}

\keywords{geometry processing, numerical coarsening, spectral geometry} 

\begin{teaserfigure}
  \centering
  \includegraphics[width=7.0in]{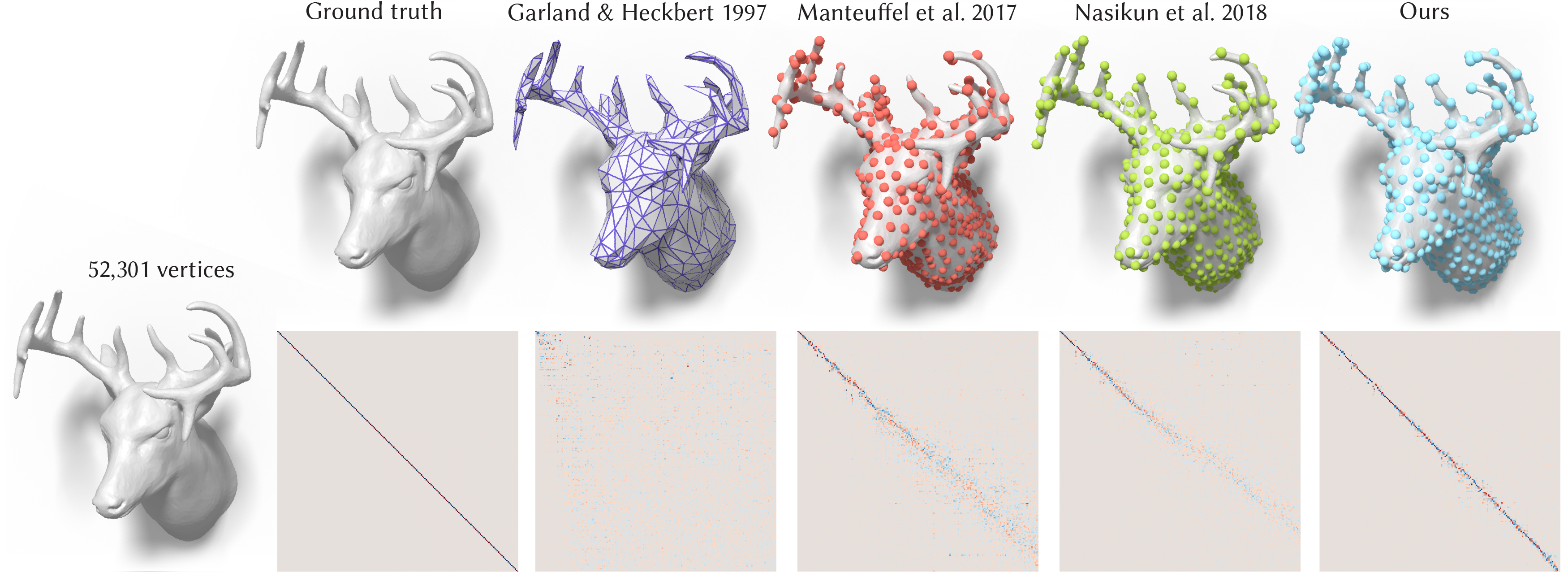}
  \caption{There are many ways to coarsen a 52,301$×$52,301 sparse anisotropic
  Laplace matrix down to a sparse 500$×$500 matrix:
  simplify the mesh \cite{garland1997surface} and rediscretize;
  apply algebraic multigrid coarsening \cite{manteuffel2017root}; or approximate
  using radial-basis functions \cite{nasikun2018fast}.
  We introduce a way to measure how well the coarse operator maintains the
  original operator's eigenvectors (bottom row).
  The visualization shows deviation from a diagonal matrix indicating poor
  eigenvector preservation.
  In response, we introduce an optimization to coarsen geometric operators while
  preserving eigenvectors and maintaining sparsity and positive
  semi-definiteness.
  }
  \label{fig:teaser}
\end{teaserfigure}

\maketitle
\section{Introduction}\label{sec:intro}

Geometry processing relies heavily on building matrices to represent linear operators defined on geometric domains.
While typically sparse, these matrices are often too large to work with efficiently when defined over high resolution \update{representations}.
A common solution is to simplify or coarsen the domain.
However, matrices built from coarse \update{representations} often do not behave the same way as their fine counterparts leading to inaccurate results and artifacts when resolution is
restored.
Quantifying and categorizing \emph{how} this behavior is different is not straightforward and most often coarsening is achieved through operator-oblivious remeshing.
The common appearance-based or geometric metrics employed by remeshers, such as the classical quadratic error metric \cite{garland1997surface} can have very little correlation to maintaining operator behavior.

We propose a novel way to compare the spectral properties of a discrete operator before and after coarsening, and to guide the coarsening to preserve them.
Our method is motivated by the recent success of spectral methods in shape analysis and processing tasks, such as shape comparison and non-rigid shape matching, symmetry detection, and vector field design to name a few. 
These methods exploit eigenfunctions of various operators, including the Laplace-Beltrami operator, whose eigenfunctions can be seen as a generalization of the Fourier basis to curved surfaces. 
Thus, spectral methods expand the powerful tools from Fourier analysis to more general domains such as shapes, represented as triangle meshes in 3D. 
We propose to measure how well the eigenvectors (and by extension eigenvalues) of a matrix $\L ∈ \R^{n×n}$ on
\begin{wrapfigure}[7]{r}{1.0in}
    \raggedleft
    \vspace{-10pt}
    \hspace*{-0.65\columnsep}
    \includegraphics[width=1.18in, trim={0mm 0mm 0mm 0mm}]{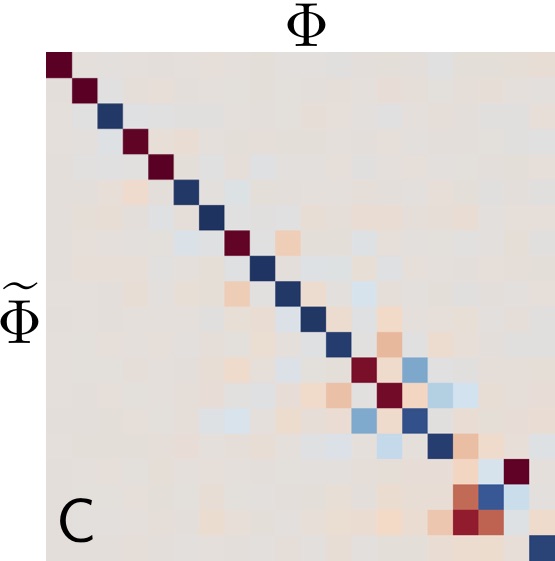} 
    \label{fig:smallFMap}
\end{wrapfigure} 
the high-resolution domain are maintained by its coarsened counterpart $\Lc ∈ \R^{m×m}$ ($m<n$) by
computing a dense matrix $\C^{k×k}$, defined as the inner product of the first $k$ eigenvectors
$\Phi∈\R^{n×k}$ and $\Phic ∈ \R^{m×k}$ of $\L$ and $\Lc$ respectively:

\vspace{0.25cm}
\noindent\begin{minipage}{1.97in}
\raggedright
\begin{equation}
  \label{equ:functional-map}
\C = \Phic^\top \Mc \P \Phi,
\end{equation}
\end{minipage}
\vspace{0.25cm}

\noindent%
where $\Mc ∈ \R^{m×m}$ defines a mass-matrix on the coarse domain and $\P∈\R^{m×n}$ is a \update{restriction} operator from fine to coarse.
The closer $\C$ resembles the identity matrix the more the eigenvectors of \update{the two operators} before and after coarsening are equivalent.
\begin{figure}
  \centering
  \includegraphics[width=3.33in]{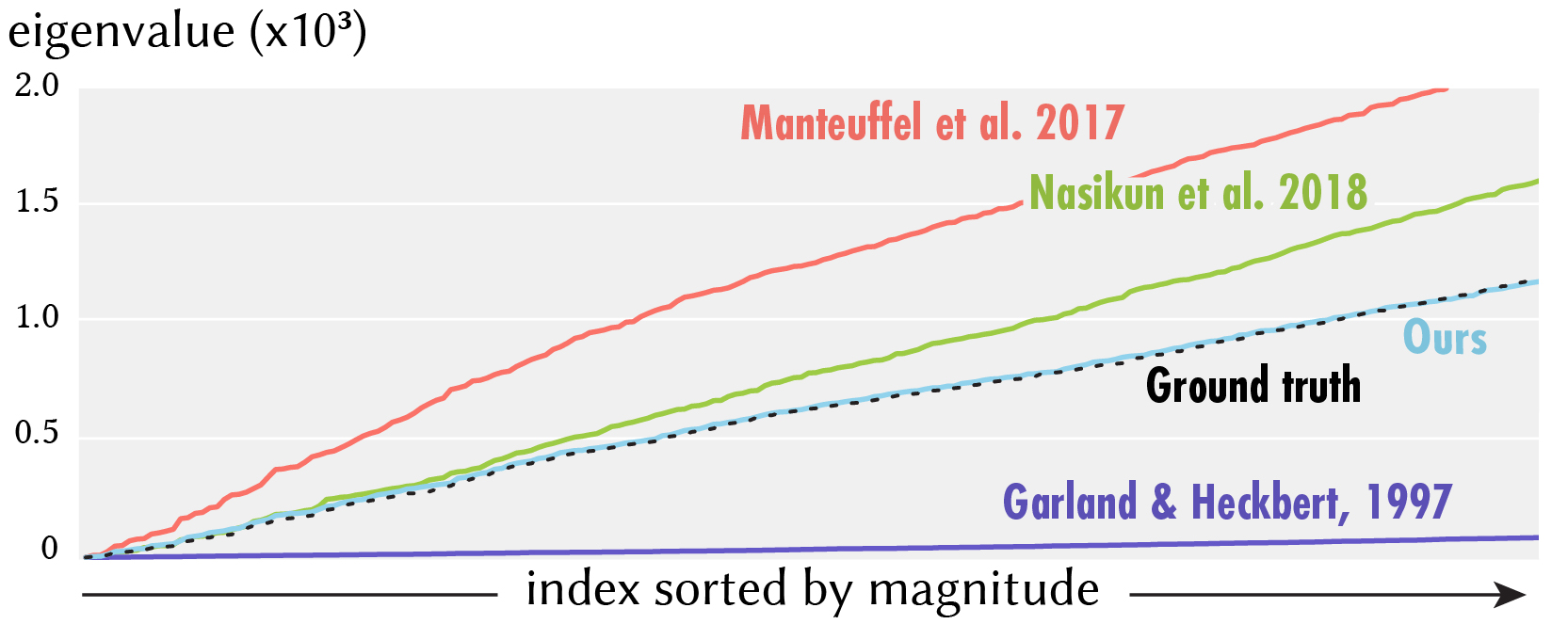}
  \caption{Our coarsening directly preserves eigenvectors so eigenvalues are \update{also implicitly} preserved: eigenvalue plot of \reffig{teaser}.
  }
  \label{fig:teaser_eVal}
  \vspace{-10pt}
\end{figure}  

We show through a variety of examples that existing geometric and algebraic coarsening methods fail
to varying degrees to preserve the eigenvectors and the eigenvalues of common operators used in geometry processing (see \reffig{teaser} and \reffig{teaser_eVal}).

In response, we propose a novel coarsening method that achieves much better preservation under this new metric.
We present an optimization strategy to coarsen an input positive semi-definite matrix in a way that better maintains its eigenvectors (see \reffig{teaser}, right) while preserving matrix sparsity and semi-definiteness.
Our optimization is designed for operators occurring in geometry processing and computer graphics,
but does not rely on access to a geometric mesh: our input is the matrix $\L$, and an area measure $\M$ on the fine domain, allowing us to deal with non-uniform sampling.
The output coarsened operator $\Lc$ and an area measure $\Mc$ on the coarse domain are defined for a subset of the input elements chosen carefully to respect anisotropy and irregular mass distribution defined by the input operator.
The coarsened operator is optimized \update{via} a novel formulation of coarsening as a sparse semi-definite programming optimization based on the operator commutative diagram.

We demonstrate the effectiveness of our method at categorizing the failure of existing methods to
maintain eigenvectors on a number of different examples of geometric domains including triangle meshes, volumetric tetrahedral meshes and point clouds.
In direct comparisons, we show examples of successful spectral coarsening for isotropic and anisotropic operators. Finally, we provide evidence that spectral coarsening can improve downstream applications such as shape matching, graph pooling for graph convolutional neural networks, and
data-driven mesh sampling.

\section{Related Work}

\begin{figure}
  \centering
  \includegraphics[width=\linewidth]{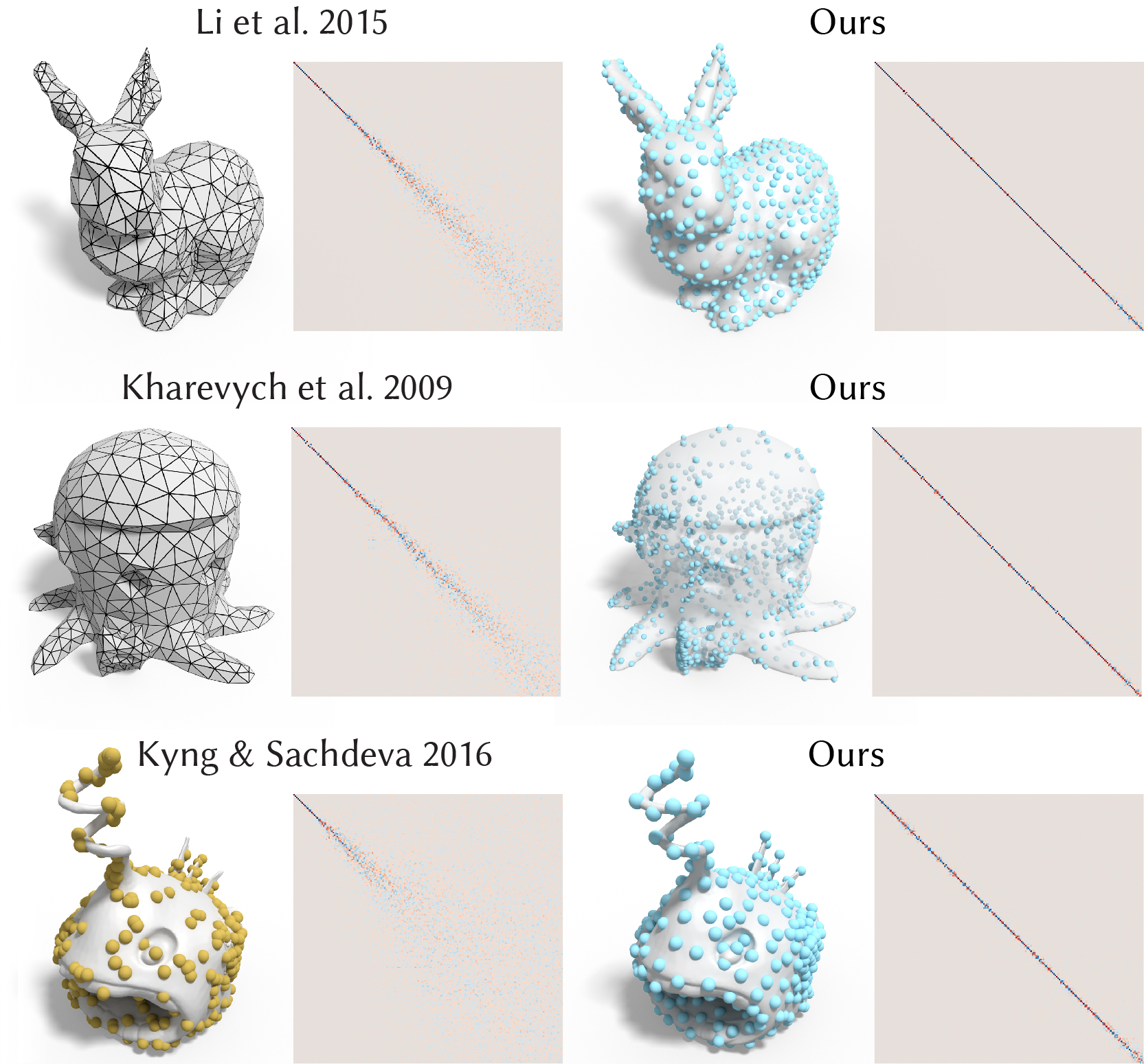}
  \caption{\update{As methods of \citet{Li:2015:transfer, kharevych2009numerical, kyng2016approximate} are not designed for preserving spectral properties, they only preserve} very low frequency
  eigenvectors (top-left corner of matrix images), but fails for subsequent modes.}
  \label{fig:comparison_related}
  \vspace{-10pt}
\end{figure}

\paragraph{Mesh Simplification and Hierarchical Representation}
The use of multi-resolution shape representations based on mesh simplification has
been extensively studied in computer graphics, with most prominent early examples including mesh
decimation and optimization approaches \cite{schroeder1992decimation,hoppe93} and their
multiresolution variants e.g., \emph{progressive meshes}
\cite{hoppe1996progressive,popovic1997progressive} (see \cite{cignoni1998comparison} for an overview
and comparison of a wide range of mesh simplification methods). Among these classical techniques,
perhaps the best-known and most widely used approach is based on the \emph{quadratic error metrics}
introduced in \cite{garland1997surface} and extended significantly in follow-up works e.g., to
incorporate texture and appearance attributes \cite{garland1998simplifying,hoppe1999new} to name a
few. Other, more recent approaches have also included variational shape approximation
\cite{cohen2004variational} and wavelet-based methods especially prominent in shape compression
\cite{schroder1996wavelets,peyre2005surface}, as well as more flexible multi-resolution approaches
such as those based on hybrid meshes \cite{guskov2002hybrid} among myriad others. Although mesh
simplification is a very well-studied problem, the vast majority of approximation techniques \update{is}
geared towards preservation of shape \emph{appearance} most often formulated via the preservation of
local geometric features.
\citet{Li:2015:transfer} conduct a \emph{frequency-adaptive} mesh simplification to better preserve the acoustic transfer of a shape by appending a modal displacement as an extra channel during progressive meshes.
In \reffig{comparison_related}, we show that this method fails to preserve all low frequency eigenvectors (since it is designed for a single frequency).
Our measure helps to reveal the accuracy of preserving \emph{spectral} quantities, and \update{to} demonstrate that existing techniques often fail to achieve this objective.

\vspace{-5pt}
\paragraph{Numerical Coarsening in Simulation}
Coarsening the geometry of an elasticity simulation mesh without adjusting the material parameters (e.g., Young's modulus) leads to \emph{numerical stiffening}.
\citet{kharevych2009numerical} recognize this and propose a method to independently adjust the per-tetrahedron elasticity tensor of a coarse mesh to agree with the
six smallest deformation modes of a fine-mesh inhomogeneous material object (see \reffig{comparison_related} for comparison).
%
\citet{Chen:2015:DFE} extend this idea via a data-driven lookup table.
\citet{Chen:2018:NCU} consider numerical coarsening for regular-grid domains, where matrix-valued basis functions on the coarse domain are optimized to again agree with the six smallest deformation modes of a fine mesh through a global
quadratic optimization.
To better capture vibrations, \citet{DAC2017} coarsen regular-grids of homogeneous materials until their low frequency vibration modes exceeding a Hausdorff distance threshold. The ratio of the first eigenvalue before and
after coarsening is then used to rescale the coarse materials Young's modulus.
In contrast to these methods, our proposed optimization is not restricted to regular grids or limited by adjusting physical parameters directly.



\begin{figure*}
  \centering
  \includegraphics[width=7.0in]{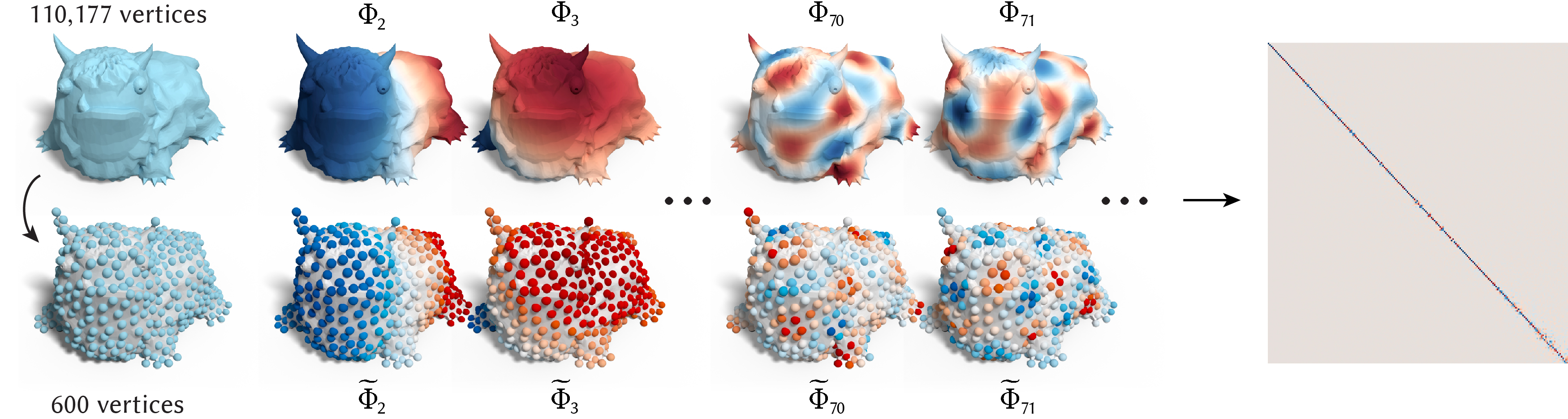}
  \caption{When eigenvectors are equivalent (up to sign) before and after
  coarsening the operator, the matrix $\C$ (right) resembles the identity
  matrix.
  }
  \label{fig:visualizeFMap}
  \vspace{-10pt}
\end{figure*}  

\vspace{-5pt}
\paragraph{Algebraic Multigrid}
Traditional multigrid methods coarsen the mesh of the geometric domain
recursively to create an efficient iterative solver for large linear systems
\cite{briggs2000multigrid}.
For isotropic operators, each geometric level smooths away error at the
corresponding frequency level \cite{burt1983laplacian}.
\emph{Algebraic} multigrid (AMG) does not see or store geometric levels, but
instead defines a hierarchy of system matrices that attempt to smooth away error
according to the input matrix's spectrum \cite{xu2017algebraic}.
AMG has been successfully applied for anisotropic problems such as cloth
simulation \cite{tamstorf2015smoothed}.
Without access to underlying geometry, AMG methods treat the input sparse matrix
as a graph with edges corresponding to non-zeros and build a coarser
graph for each level by removing nodes and adding edges according to an
\emph{algebraic distance} determined by the input matrix.
AMG like all multigrid hierarchies are typically measured according to their
solver convergence rates \cite{xu2017algebraic}.
While eigenvector preservation is beneficial to AMG, an efficient solver must
also avoid adding too many new edges during
coarsening (i.e., \cite{livne2012lean, kahl2018least}). 
Meanwhile, to remain competitive to other blackbox solvers, AMG methods also
strive to achieve very fast hierarchy construction performance
\cite{xu2017algebraic}. 
Our analysis shows how state-of-the-art AMG coarsening methods such as \cite{manteuffel2017root} \update{which is designed for fast convergence}
fails to preserve eigenvectors and eigenvalues (see \reffig{teaser} and \reffig{teaser_eVal}).
Our subsequent optimization formulation in \refsec{coarsening} and
\refsec{operator-optimization} is inspired by the ``root node'' selection and
Galerkin projection approaches found in the AMG literature
\cite{stuben2000algebraic,bell2008algebraic,manteuffel2017root}.

\vspace{-5pt}
\paragraph{Spectrum Preservation}
In contrast to geometry-based mesh simplification very few methods have been
proposed targeting preservation of spectral properties. {\"O}ztireli and
colleagues \cite{oztireli2010spectral} introduced a technique for spectral
sampling on surfaces. In a similar spirit to our approach, their method aims to
compute samples on a surface that can approximate the Laplacian spectrum of the
original shape. This method targets only isotropic sampling and is not
well-suited to more diverse operators such as the anisotropic Laplace-Beltrami
operator handled by our approach. More fundamentally, our goal is to construct a
coarse representation that preserves an entire \emph{operator}, and allows, for
example to compute eigenfunctions and eigenvalues in the coarse domain, which is
not addressed by a purely sampling-based strategy. More recently, an efficient
approach for approximating the Laplace-Beltrami eigenfunctions has been
introduced in \cite{nasikun2018fast}, based on a combination of fast Poisson
sampling and an adapted coarsening strategy. While very efficient, as we show
below, this method unfortunately fails to preserve even medium frequencies,
especially in the presence of high-curvature shape features or more diverse,
including anisotropic Laplacian, operators.

We note briefly that spectrum preservation and optimization has also been considered in the context of sound synthesis, including \cite{bharaj2015computational}, and more algebraically for efficient solutions of Laplacian linear systems \cite{kyng2016approximate}. 
In \reffig{comparison_related}, we show that the method of \citet{kyng2016approximate} only preserves very low frequency eigenvectors. 
Our approach is geared towards operators defined on non-uniform triangle meshes and does not have
limitations of the approach of \cite{kyng2016approximate} which only works on Laplacians where
all weights are positive.

\section{Method}
\label{sec:method}
The input to our method is a $n$-by-$n$ sparse, positive semi-definite matrix $\L ∈ \R^{n×n}$.
We assume $\L$ is the Hessian of an energy derived from a geometric domain with
$n$ vertices and the sparsity pattern is determined by the connectivity of a
mesh or local neighbor relationship.
For example, $\L$ may be the discrete cotangent
Laplacian, the Hessian of the discrete Dirichlet energy. 
However, we do not require direct access to the geometric domain or its spatial
embedding.
We also take as input a non-negative diagonal weighting or mass matrix $\M ∈ \R^{n×n}$ (i.e., defining the inner-product space of vectors from the input
domain).
The main parameter of our method is the positive number $m < n$ which determines
the size of our coarsened output.

Our method outputs a sparse, positive semi-definite matrix $\Lc ∈ \R^{m×m}$ that
attempts to maintain the low-frequency eigenvalues and eigenvectors of the input
matrix $\L$ (see \reffig{visualizeFMap}).
\vspace{-5pt}
\begin{algorithm}
  \caption{Spectral Coarsening given $\L$, $\M$ and $m$}
    $\P,\update{\K} ← \textit{combinatorial coarsening}(\L,\M,m)$; \\
    $\Lc,\Mc ← \textit{operator optimization}(\L,\M,\P,\update{\K})$;
  \label{alg:method}
\end{algorithm} 
\vspace{-5pt}

We propose coarsening in two steps (see \refalg{method}). First we treat the input matrix $\L$ as encoding a graph and select a subset of $m$ ``root'' nodes, assigning all others
to clusters based on a novel graph-distance. This clustering step defines a \update{restriction} operator ($\P$ in \refequ{functional-map}) and \update{a cluster-assignment operator $\K$} that determines the sparsity pattern of our output matrix $\Lc$. In the second stage, we optimize the non-zero values of $\Lc$.

\subsection{Combinatorial coarsening} \label{sec:coarsening}
\begin{wrapfigure}[13]{r}{0.4\linewidth}
  \vspace*{-\intextsep}
  \hspace*{-0.5\columnsep}
  \begin{minipage}[b]{\linewidth}
  \includegraphics[width=\linewidth, trim={0mm 3mm 0mm 0mm}]{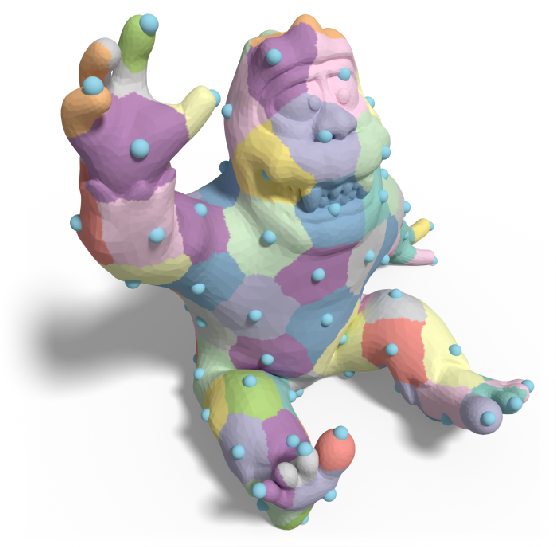}
  \caption{\label{fig:clusters} Blue dots and colored regions indicate ``root nodes'' and clusters selected by
  $\P$ and $\K$ respectively.}
  \end{minipage}
\end{wrapfigure}

Given an input operator $\L ∈ \R^{n×n}$ and corresponding mass-matrix $\M∈\R^{n×n}$, the goal of this stage is to construct two sparse binary matrices $\K,\P ∈ \{0,1\}^{m×n}$ (see \reffig{clusters}). 
Acting as a cluster-assignment operator, \update{$\K$ has exactly one $1$ per column,}
so that $\K_{ij} = 1$ indicates that element
$j$ on the input domain is assigned to element $i$ on the coarsened domain.
Complementarily, acting as a \update{restriction} or subset-selection operator, \update{$\P$ has exactly one $1$ per row and} 
\update{no more than one $1$ per column,}
%
so that $\P_{ij} = 1$ indicates that element $j$ on the input domain is selected as \update{element $i$ in the coarsened domain to \emph{represent} its corresponding cluster}.
Following the terminology from the algebraic multigrid literature, we refer to this selected element as the ``root node'' of the cluster \cite{manteuffel2017root}. 
In our figures, we visualize $\P$ by drawing large dots on 
the selected nodes \update{and $\K$ by different color segments.}

\begin{figure}
  \centering
  \includegraphics[width=3.33in]{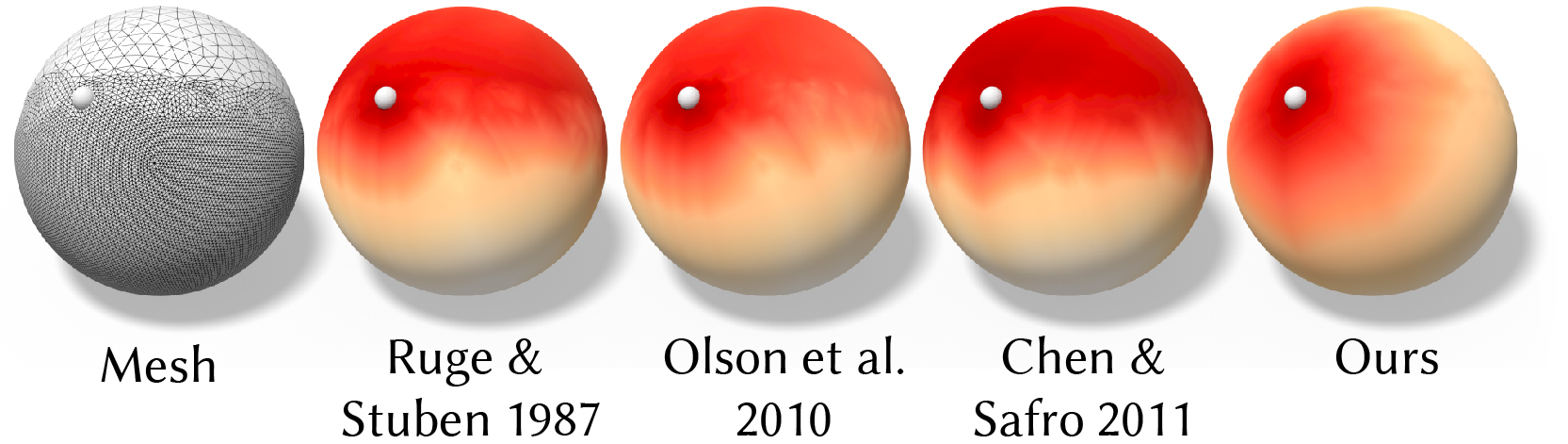}
  \caption{We visualize the graph shortest path distance from the source point (gray) to all the other points, where the strength of connections between adjacent points is defined using different operator-dependent strength measures. In an isotropic problem, our resulting ``distance'' is more robust to different element sizes and grows more uniformly in all directions (right).}
  \label{fig:algebraicDistances}
  \vspace{-5pt}
\end{figure}

Consider the graph with $n$ nodes implied by interpreting non-zeros of $\L$ as undirected edges. Our node-clustering and root-node selection should respect how quickly information at one node \emph{diffuses} to neighboring nodes
according to $\L$ and how much mass or weight is associated with each node according to $\M$.
Although a variety \update{of} algebraic distances have been proposed
\cite{ruge1987algebraic, chen2011algebraic, olson2010new,
livne2012lean}, they are not directly applicable to our geometric tasks because they are sensitive to different finite-element sizes (see \reffig{algebraicDistances}). 

According to this diffusion perspective, the edge-distance of the edge between nodes $i$ and $j$ should be inversely correlated \update{with} $-\L_{ij}$ and positively correlated \update{with $(\M_{ii}+\M_{jj})$}. Given the \emph{units} of $\L$ and $\M$ in terms of powers of length $p$ and $q$ respectively (e.g., the discrete cotangent Laplacian for a triangle mesh has \update{units $p$=0}, the barycentric mass matrix has \update{units $q$=2}), then we adjust these correlations so that our edge-distance has units of length. 
Putting these relations together and avoiding negative lengths due to positive off-diagonal entries in $\L$, we define the edge-distance between connected nodes as:
\update{
\begin{align*}
  \D_{ij} = \max\left(\frac{(\M_{ii} + \M_{jj})^{(p+1)/q}}{-\L_{ij}},0\right).
\end{align*}
}

\begin{wrapfigure}[11]{r}{0.45\linewidth}
  \vspace*{-0.9\intextsep}
  \hspace*{-0.5\columnsep}
  \begin{minipage}[b]{1.05\linewidth}
  \includegraphics[width=1.05\linewidth,  trim={0mm 4mm 0mm 0mm}]{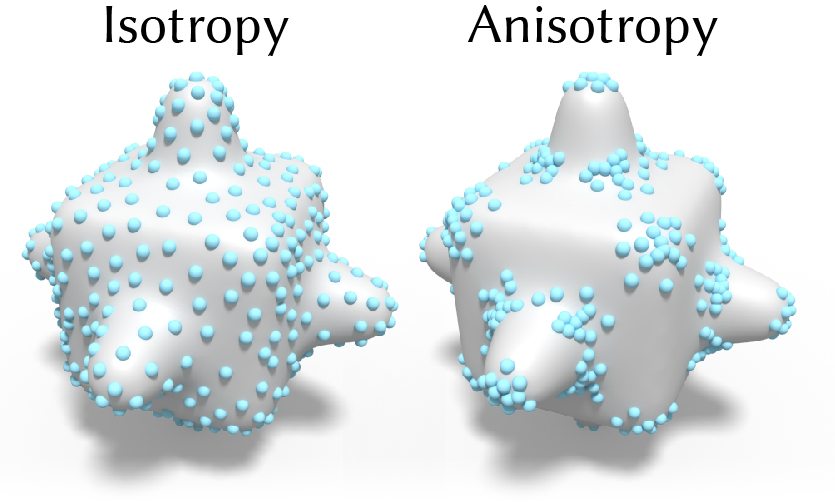}
  \caption{\label{fig:anisoBumpy} Our coarsening is aware of the anisotropy of the underlying operator, resulting in a different set of selected root nodes.}
  \end{minipage}
\end{wrapfigure}
\begin{figure}
  \centering
  \includegraphics[width=3.33in]{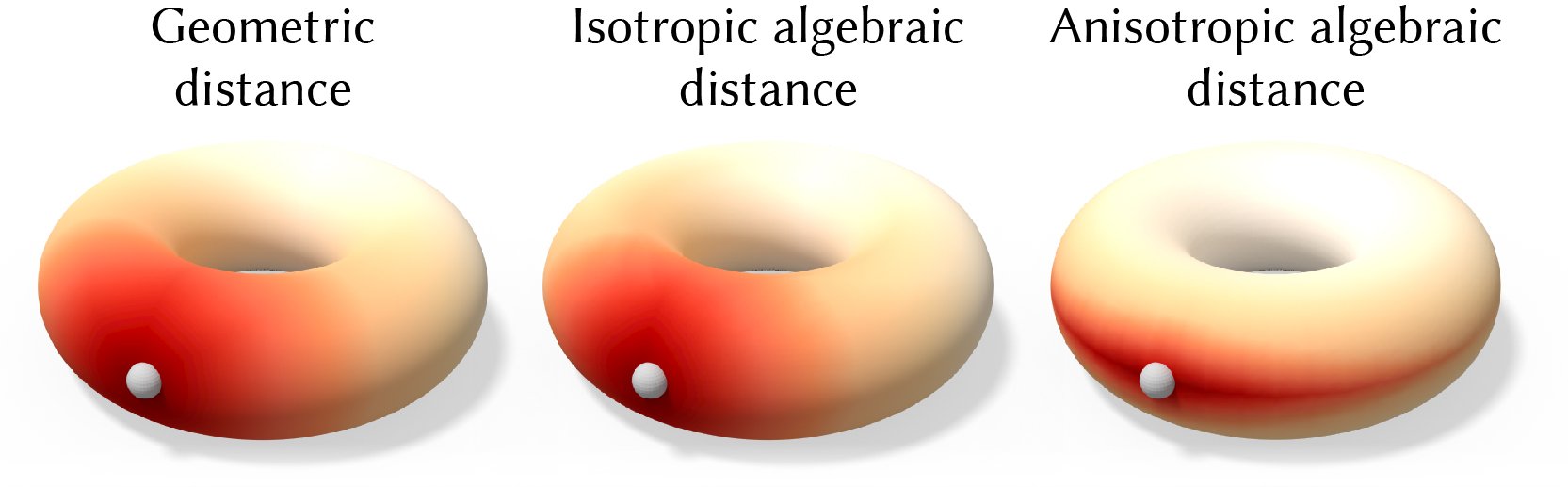}
  \caption{We visualize the graph shortest path distance from the source point (gray) to all the other points. \update{Our} operator-dependent distance can handle both isotropic and anisotropic problems, \update{whereas standard} geometry-based measure (e.g. edge
  length) is limited to isotropic problems.}
  \label{fig:geoAlgDist}
  \vspace{-5pt}
\end{figure}

Compared to Euclidean or geodesic distance, shortest-path distances using this edge-distance will respect anisotropy of $\L$ (see \reffig{anisoBumpy}, \reffig{geoAlgDist}).
Compared to state-of-the-art \update{algebraic distances}, our distance \update{will account} for irregular mass distribution, e.g., due to irregular meshing (see \reffig{algebraicDistances}).

\begin{figure}
  \centering
  \includegraphics[width=3.33in]{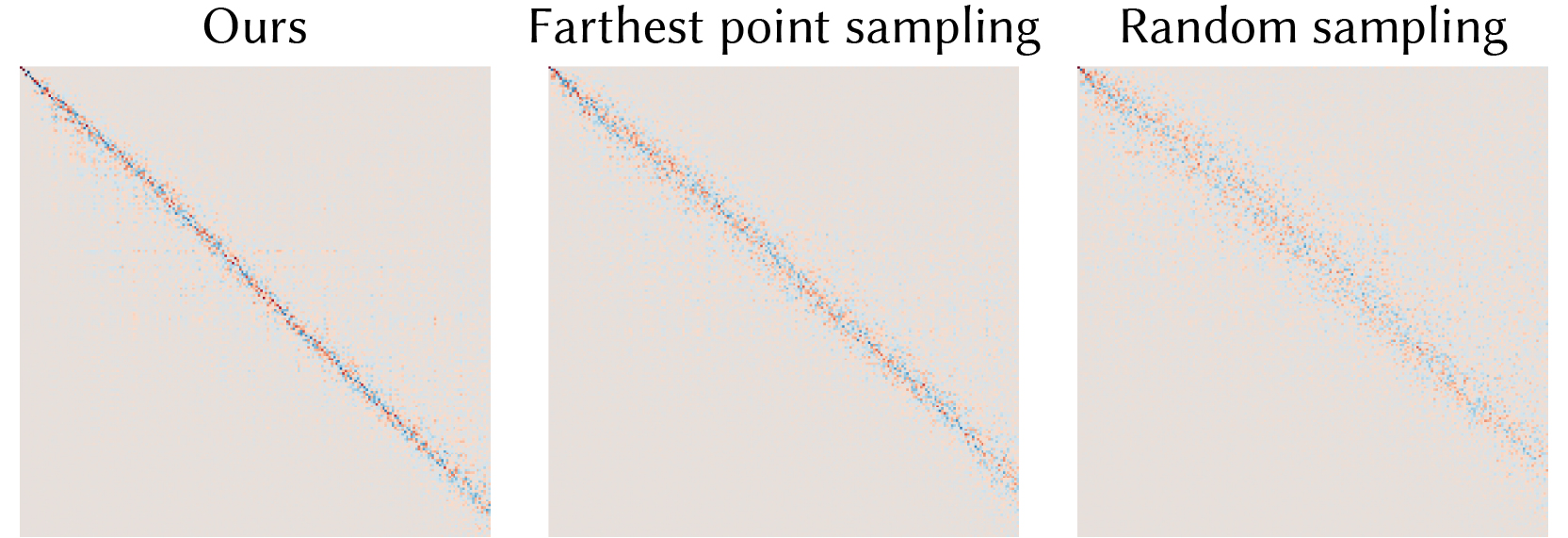}
  \caption{\update{Coarsening with our operator-aware distance (left) results in a better eigenfunction preservation compared to the farthest point sampling (middle) and the random sampling (right) on an anisotropic operator.}}
  \label{fig:sampling}
  \vspace{-5pt}
\end{figure}

Given this (symmetric) matrix of edge-distances, we compute the $k$-mediods clustering \cite{JSSv001i04} of the graph nodes according to shortest path distances (computed efficiently using the modified Bellman-Ford method and Lloyd
aggregation method of \citet{bell2008algebraic}).
We initialize this iterative optimization with a random set of $k$ root nodes.
Unlike $k$-means where the \emph{mean} of each cluster \update{is not restricted to the set of input points} in space, $k$-mediods chooses the cluster root as the mediod-node of the cluster (i.e., the node with minimal total distance to all other nodes in the cluster).
All other nodes are then re-assigned to their closest root. This process is iterated until convergence. Cluster assignments and cluster roots are stored as \update{$\K$} and \update{$\P$} accordingly.
\update{Comparing to the farthest point sampling and the random sampling, our approach results in a better eigenfunction preservation image for anisotropic operators (\reffig{sampling}).}

We construct a sparsity pattern for $\Lc$ so that $\Lc_{ij}$ may be non-zero if the cluster $j$ is in the three-ring \update{neighborhood} of cluster $i$ as determined by cluster-cluster adjacency.
If we let $\SL ∈ \{0,1\}^{n×n}$ be a binary matrix containing a 1 if and only if the corresponding element of $\L$ is non-zero, then we can compute the ``cluster adjacency'' matrix $\Ac = \K \SL \K^⊤ ∈ \{0,1\}^{m×m}$ so that $\Ac_{ij} = 1$
if and only if the clusters $i$ and $j$ contain some elements $u$ and $v$ such
that $\L_{uv} ≠ 0$. \update{Using this adjacency matrix, we create a sparse restriction matrix with wider connectivity $\SG = \K^⊤ \Ac ∈ \{0,1\}^{n×m}$.}
Finally, our predetermined sparsity pattern for $\Lc$ is defined to be that of
$\SLc = \SG^\top \SL \SG = \Ac³ ∈ \{0,1\}^{m×m}$.
We found that using the cluster three-ring sparsity is a reasonable trade-off between in-fill density and performance of the optimized operator. \update{Assuming the cluster graph is 2-manifold with average valence 6, the three-ring sparsity implies that $\Lc$ will have 37 non-zeros per row/column on average, independent to $m$ and $n$. In practice, our cluster graph is nearly 2-manifold. The $\Lc$ in \reffig{clusters}, for instance, has approximately $39$ non-zeros per row/column.}

\subsection{Operator optimization} \label{sec:operator-optimization}
Given a clustering, root node selection and \update{the} desired sparsity pattern, our second step is to compute a coarsened matrix $\Lc$ that maintains the eigenvectors of the input matrix $\L$ as much as possible. 
Since $\L$ and $\Lc$ are \update{of} different sizes, their corresponding eigenvectors are also \update{of} different lengths.
To compare them in a meaningful way we will use the functional map matrix $\C$ defined in \refequ{functional-map} implied by the \update{restriction} operator $\P$ \update{(note that: prolongation from coarse to fine is generally ill-defined)}. This also requires a mass-matrix on the coarsened domain, which we compute by lumping cluster masses: $\Mc = \K \M \K^⊤$.
The first $k$ eigenvectors for the input operator and yet-unknown coarsened operator may be computed as solutions to the generalized eigenvalue problems \update{$\L \Phi = Λ \M \Phi $ and $\Lc \Phic = \coarse{Λ} \Mc \Phic $, where $Λ, \coarse{Λ}$ are eigenvalue matrices.}

\begin{wrapfigure}[10]{r}{0.45\linewidth}
  \vspace*{-0.6\intextsep}
  \hspace*{-0.4\columnsep}
  \begin{minipage}[b]{1.05\linewidth}
  \includegraphics[width=1.02\linewidth,  trim={3mm 2mm 0mm 0mm}]{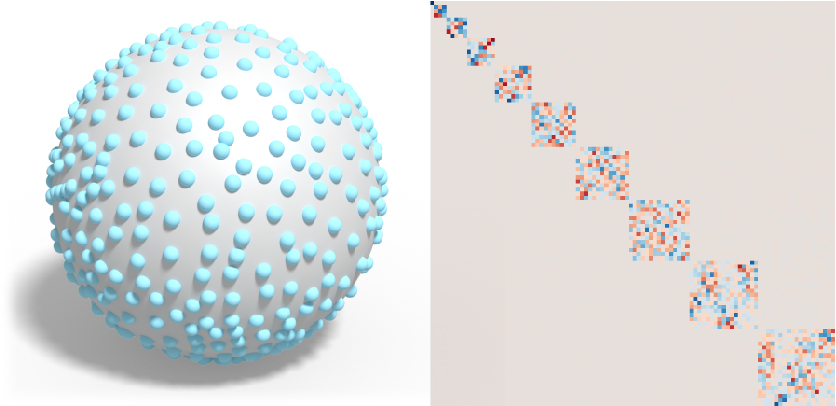}
  \caption{\label{fig:multiplicity} The optimized $\C$ should be block diagonal when the operator has algebraic multiplicity.}
  \end{minipage}
\end{wrapfigure}
Knowing that the proximity of the functional map matrix $\C$ to an identity matrix encodes eigenvector preservation, it might be tempting to try to enforce \update{$\|\C - \I\|_F$ directly}.
This however is problematic because it does not handle sign flips or multiplicity (see \reffig{multiplicity}).
More importantly, recall that in our setting $\C$ is not a free variable, but rather a non-linear function (via eigen decomposition) of the unknown sparse matrix $\Lc$. 

Instead, we propose to minimize \update{the} failure to realize the commutative diagram of a functional map.
Ideally, for any function on the input domain $\f ∈ \R^n$ applying the input operator $\M^{-1}\L$ and then the \update{restriction} matrix $\P$ is equivalent to applying $\P$ then $\Mc^{-1}\Lc$, resulting in the same function $\coarse{\f} ∈ \R^m$ on the coarsened domain: 
\begin{align}
  \label{equ:commutative-diagram}
  \begin{array}{rclcccc}
         & \M^{-1}\L                   &   \\
      {\f\hspace{0.18em}} & \xrightarrow{\hspace*{1.2cm}} & {\hspace{0.1em}\bullet} \vspace{0.15cm}\\ 
   \P\left↓\vphantom{\int}\right. &                             &
     \left↓\vphantom{\int}\right. \P \vspace{0.15cm}\\ 
       {\bullet\hspace{0.19em}} & \xrightarrow{\hspace*{1.2cm}} & \coarse{\hspace{0.1em}\f}\\
         & \Mc^{-1}\Lc 
  \end{array}
\end{align}
This leads to a straightforward energy that minimizes \update{the difference between the two paths in} the commutative diagram for all possible functions $\f$:
\begin{align}
\label{equ:energy}
  E(\Lc) = \|\P \M^{-1} \L \I - \Mc^{-1} \Lc \P \I \|^2_{\Mc},
\end{align}
where $\I ∈ \R^{n×n}$ is the identity matrix (included didactically for the discussion that follows) and 
\begin{equation*}
  \|\X\|^2_{\Mc} =
\mathop{\text{tr}}(\X^⊤ \Mc\X)
\end{equation*}
computes the Frobenius inner-product defined 
by $\Mc$.

By using $\I$ as the spanning matrix, we treat all functions equally in an L2
sense.
Inspired by the functional maps literature, we can instead compute this energy
over only lower frequency functions spanned by the first $k$ eigenvectors $\Phi∈\R^{n×k}$ of the operator $\L$. Since high frequency functions naturally cannot
live on a coarsened 
\begin{wrapfigure}[5]{r}{1.37in}
  \raggedleft
  \vspace{-10pt}
  \hspace*{-0.7\columnsep}
  \includegraphics[width=1.31in, trim={06mm 0mm 0mm 0mm}]{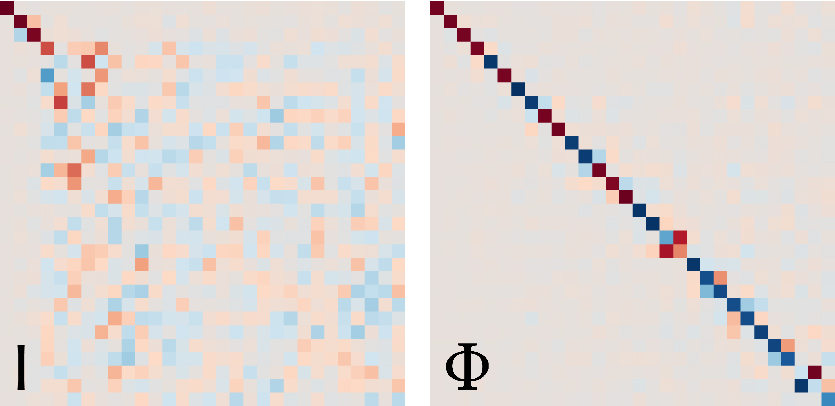} 
  \label{fig:minIdentity}
\end{wrapfigure} 
domain, this parameter $k$ allows the optimization to focus on functions that
matter. Consequently, preservation of low frequency eigenvectors dramatically
improves (see inset).

Substituting $\Phi$ for $\I$ in \refequ{energy}, we now consider minimizing this
reduced energy $E_k$ over all possible \emph{sparse} positive semi-definite
(PSD) matrices $\Lc$:
\begin{align}
  \label{equ:sdp}
  \mathop{\text{minimize}}_{\Lc ≐ \SLc}\ & \underbrace{\frac{1}{2}\|\P \M^{-1} \L \Phi - \Mc^{-1} \Lc
  \P \Phi \|^2_{\Mc}}_{E_k(\Lc)} \\
  \text{subject to }&\ \Lc \text{ is positive semi-definite} \\
  \label{equ:null-space}
  \text{and }       &\ \Lc \P \Phi_0 = 0
\end{align}
%
\update{where we use $\mat{X} ≐ \mat{Y}$ to denote that $\mat{X}$ has the same sparsity pattern as $\mat{Y}$.}
The final linear-equality constraint in \refequ{null-space} ensures that the eigen-vectors $\Phi_0$ corresponding to
zero eigenvalues are exactly preserved ($\P \L \Phi_0 = \P \M \Phi_0 0 = 0$). Note that while it might seem that
\refequ{sdp} is only meant to preserve the eigen\emph{vectors}, a straightforward calculation (See
\refapp{eValPreservation}) shows that it promotes the preservation of eigen\emph{values} as well.

This optimization problem is convex \cite{boyd2004convex}, but the sparsity
constraint makes it challenging to solve efficiently. Most efficient
semi-definite programming (SDP) solvers (e.g., \textsc{Mosek}, \textsc{cvxopt},
\textsc{Gurobi}) only implement \emph{dense} PSD constraints.  The academic
community has studied SDPs over sparse matrices, yet solutions are not
immediately applicable (e.g., those based on chordal sparsity
\cite{Vandenberghe2015CGS,ZhengFPGW17}) or practically efficient (e.g.,
\cite{Andersen2010}). Even projecting a sparse matrix $\X$ on to the set of PSD
matrices with the same sparsity pattern is a difficult sub-problem (the
so-called sparse matrix nearness problem, e.g., \cite{sun_2015}), so that
proximal methods such as ADMM lose their attractiveness.

\begin{figure}
  \centering
  \includegraphics[width=3.33in]{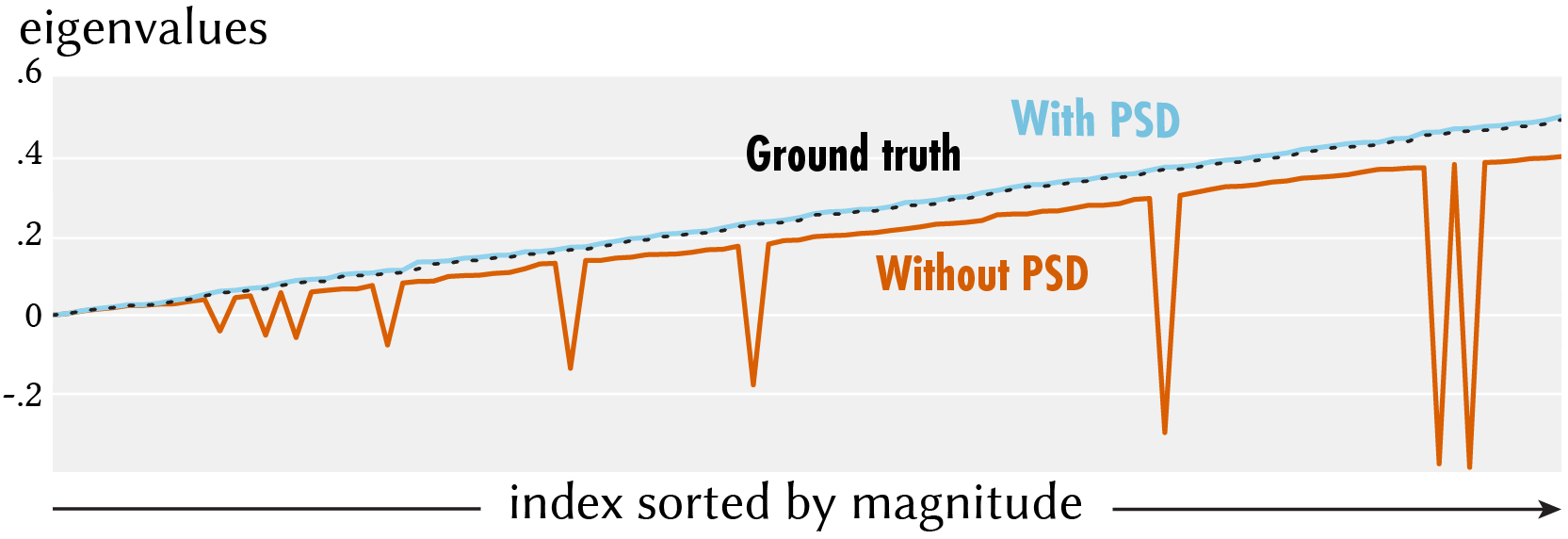}
  \caption{Dropping the PSD constraint leads to a simple quadratic optimization problem which can be solved directly, but it produces a non-PSD $\Lc$ that contains negative eigenvalues.}
  \label{fig:directSolverEVal}
  \vspace{-10pt}
\end{figure}

If we drop the PSD constraint, the result is a simple quadratic optimization
with linear constraints and can be solved directly. While this produces
solutions with very low objective values $E_k$, the eigenvector preservation is
sporadic and negative eigenvalues appear (see \reffig{directSolverEVal}).
Conversely, attempting to replace the PSD constraint with the \emph{stricter}
but more amenable diagonal dominance linear inequality constraint (i.e.,
$\Lc_{ii} ≥ ∑_{j≠i} \Lc_{ji}$) produces a worse objective value and poor
eigenvector preservation.

Instead, we propose introducing an auxiliary sparse matrix variable $\G∈\R^{n×m}$
and restricting the coarsened operator to be created by using $\G$ as an
interpolation operator: $\Lc := \G^⊤ \L \G$. Substituting this into 
\refequ{sdp}, we optimize 
\begin{align}
\label{equ:quartic}
  \mathop{\text{minimize}}_{\G ≐ \SG}\ & \underbrace{\frac{1}{2}\|\P
  \M^{-1} \L \Phi - \Mc^{-1} \G^⊤  \L \G \P \Phi
  \|^2_{\Mc}}_{E_k(\G)}, \\ \nonumber
  \text{subject to }        &\ \G \P \Phi_0 = \Phi_0
\end{align}
where the sparsity of $\Lc$ is maintained by requiring sparsity of $\G$.
The null-space constraint remains linear because $\G \P \Phi_0 = \Phi_0 ⇒
\G^⊤ \L \G \P \Phi_0  = 0$ \update{implies that $\Lc$ contains the null-space of $\L$}. \update{The converse  
will not necessarily be true}, but is unlikely to happen
because this would represent inefficient minimization of the objective. In practice, we never found that spurious null-spaces occurred.
While we get to remove the PSD constraint ($\L ≽ 0$ implies $\G^⊤ \L\G ≽ 0$), the price we have paid is that the energy is no longer
quadratic in the unknowns, but \update{quartic}.

\update{Therefore} in lieu of convex programming, we optimize this energy over the non-zeros of
$\G$ using a gradient-based algorithm with a fixed step size $γ$. \update{Specifically, we use \nadam \cite{dozat2016incorporating} optimizer which is a variant of gradient descent that combines momentum and Nesterov's acceleration.}
For completeness, we provide the \emph{sparse} matrix-valued gradient $∂E_k/∂\G$ in \refapp{dEdG}. 
The sparse linear equality constraints are handled with the orthogonal projection \update{in \refapp{sparseProjection})}. 
We summarize our optimization in pseudocode \refalg{PGD}. We stop
the optimization if it stalls (i.e., does not decrease the objective after 10
iterations) and use a fixed step size $γ = 0.02$.
This rather straightforward application of a gradient-based optimization to
maintaining the commutative diagram in \refequ{commutative-diagram} performs
quite well for a variety of domains and operators.
\vspace{-5pt}
\begin{algorithm}[h!]
  \caption{Operator optimization using \nadam}
  \update{$\G ← \K^⊤$;} \hfill \textit{\color{myGray}// initialization}\\
  \While{not stalled}{
    $\nicefrac{∂ E_k}{∂ \G} ← \textit{sparse gradient }(\G)$; \\
    $∆\G                    ← \nadam(\nicefrac{∂ E_k}{∂ \G})$;\\
    $\G                     ← \G - \gamma\ ∆\G$;\\
    \update{$\G             ← \textit{orthogonal projection}\,(\G, \Phi_0)$;} \hfill \textit{\color{myGray}// see \refapp{sparseProjection}}
  }
  \label{alg:PGD}
\end{algorithm} 
\vspace{-5 pt}
\vspace{-5pt}
\section{Evaluation \& Validation}\label{sec:evaluation}
\begin{wrapfigure}[13]{r}{0.45\linewidth}
  \vspace*{-1.0\intextsep} 
  \hspace*{-0.6\columnsep}
  \begin{minipage}[b]{1.1\linewidth}
  \includegraphics[width=1.05\linewidth, trim={1mm 2mm 0mm 0mm}]{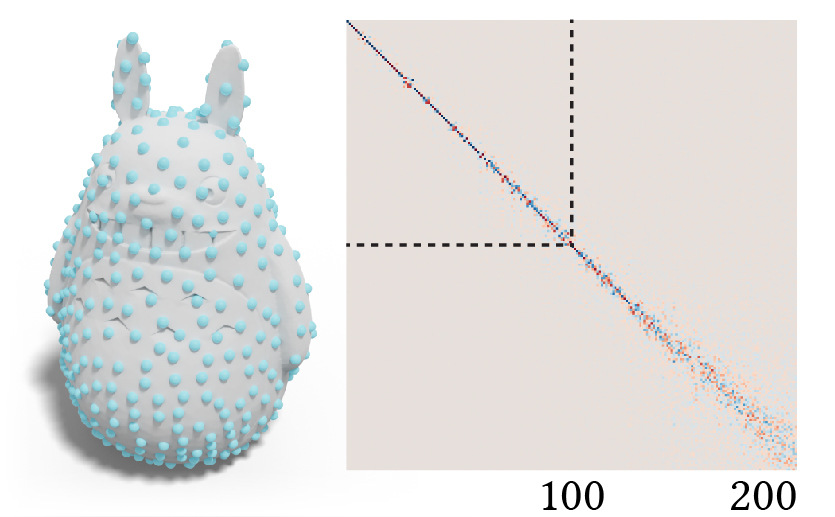}
  \caption{\label{fig:generalization} We optimize for the first 100 eigenfunctions and visualize the 200$×$200 functional map, demonstrating a graceful generalization beyond the optimized eigenfunctions.}
  \end{minipage}
\end{wrapfigure}
\begin{figure}
  \centering
  \includegraphics[width=3.33in]{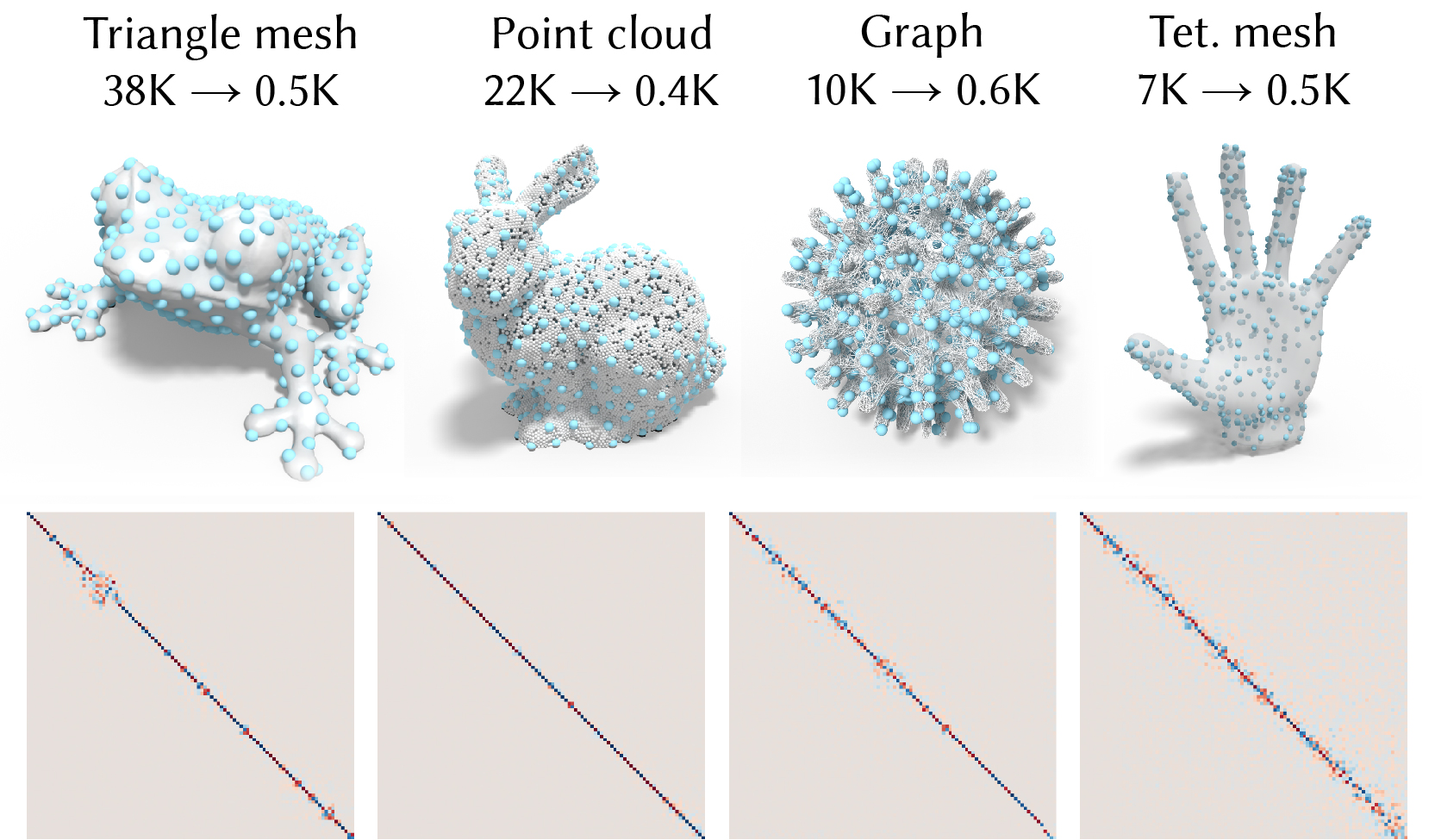}
  \caption{Our algebraic formulation is directly applicable to different data types, such as triangle meshes, point clouds, graphs, and tetrahedral meshes.}
  \label{fig:difDomains}
  \vspace{-10pt}
\end{figure}
\begin{figure}
  \centering
  \includegraphics[width=3.33in]{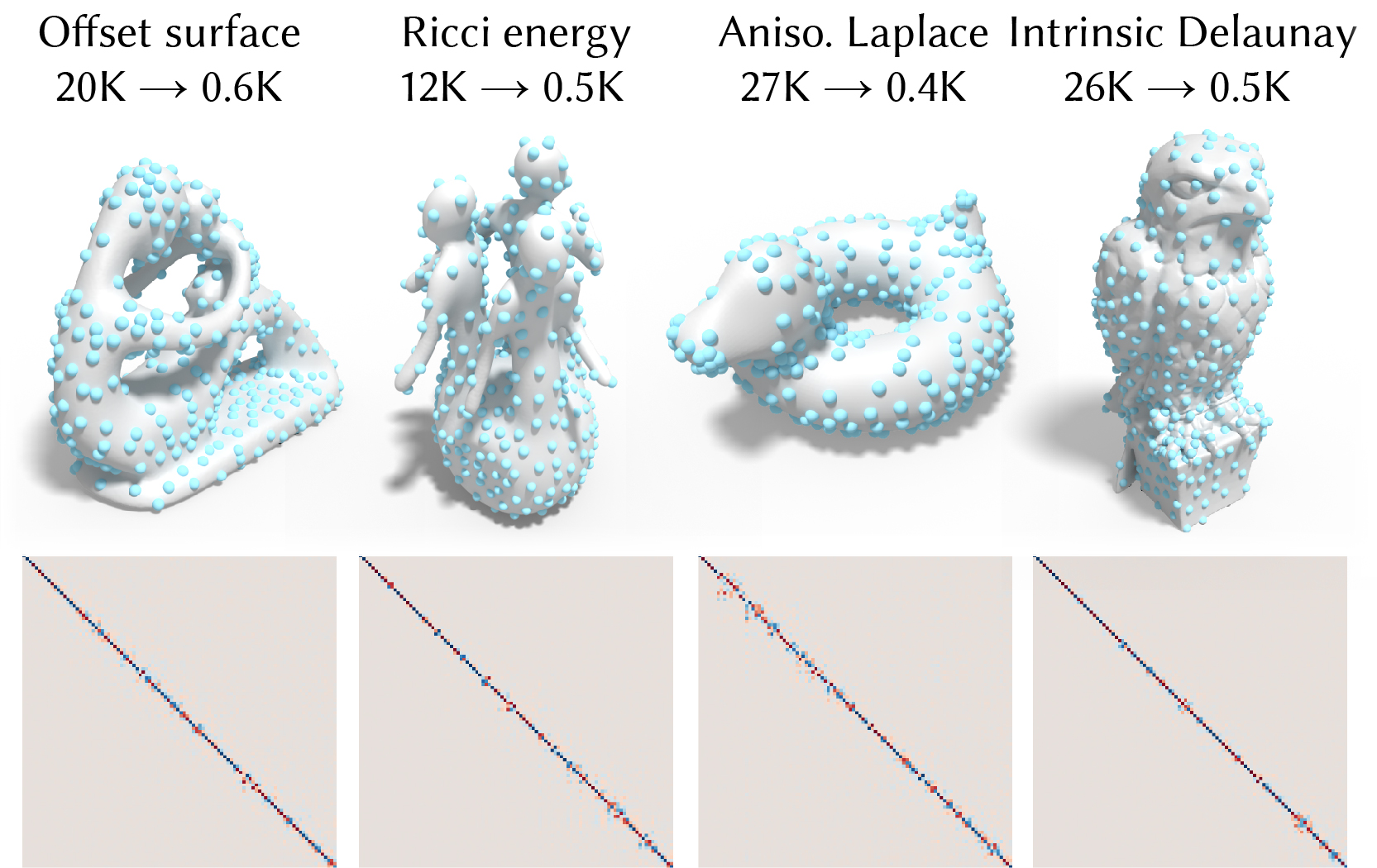}
  \caption{Our method preserves the eigenfunctions of the offset surface Laplacian, the Hessian of the Ricci energy, the anisotropic Laplace, and the the intrinsic Delaunay Laplacian.}
  \label{fig:difOpts}
  \vspace{-10pt}
\end{figure}

Our input is a matrix $\L$ which can be derived from a variety of geometric data types. In
\reffig{difDomains} we show that our method can preserve the property of the Laplace operators
defined on triangle meshes \cite{pinkall1993computing, macneal1949solution, desbrun1999implicit},
point clouds \cite{belkin2009constructing}, graphs, and tetrahedral meshes \cite{sharf2007interactive}.
We also evaluate our method on a variety of operators, including the offset
surface Laplacian \cite{corman2017functional}, the Hessian of the Ricci energy \cite{jin2008discrete}, anisotropic Laplacian \cite{andreux2014anisotropic}, and the intrinsic Delaunay Laplacian \cite{fisher2006algorithm} (see \reffig{difOpts}).

\update{We further evaluate how the coarsening generalizes beyond the optimized eigenfunctions. In \reffig{generalization}, we coarsen the shape using the first 100 eigenfunctions and visualize the 200$×$200 functional map image. This shows a strong diagonal for the upper 100$×$100 block and a slowly blurring off-diagonal for the bottom block, demonstrating a graceful generalization beyond the optimized eigenfunctions.}

\begin{figure}
  \centering
  \includegraphics[width=3.33in]{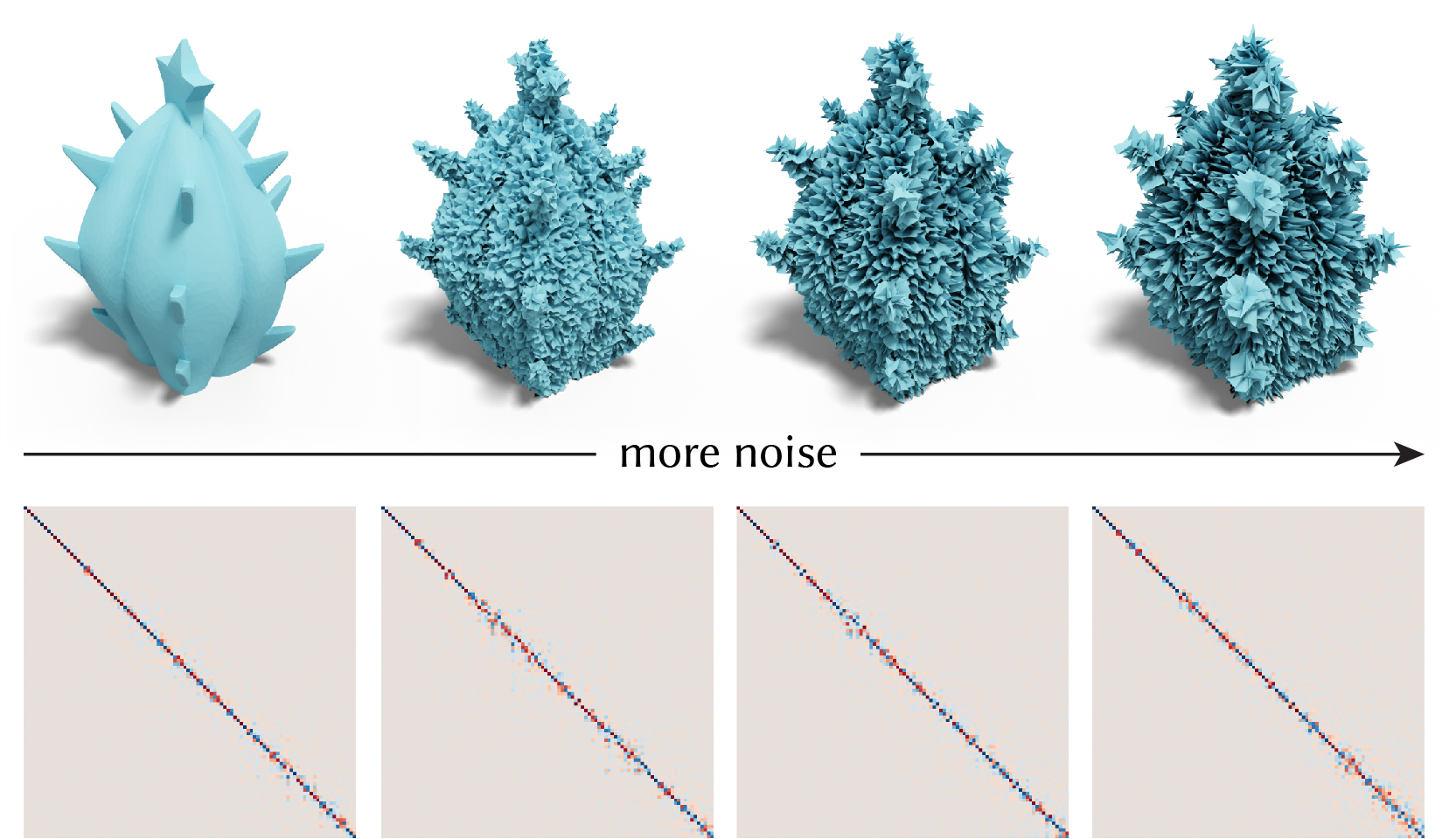}
  \caption{\update{Our coarsening takes the operator as the input, thus the output quality is robust to noise and sharp geometric features.}}
  \label{fig:meshNoise}
  \vspace{-10pt}
\end{figure}

\begin{figure}
  \centering
  \includegraphics[width=3.33in]{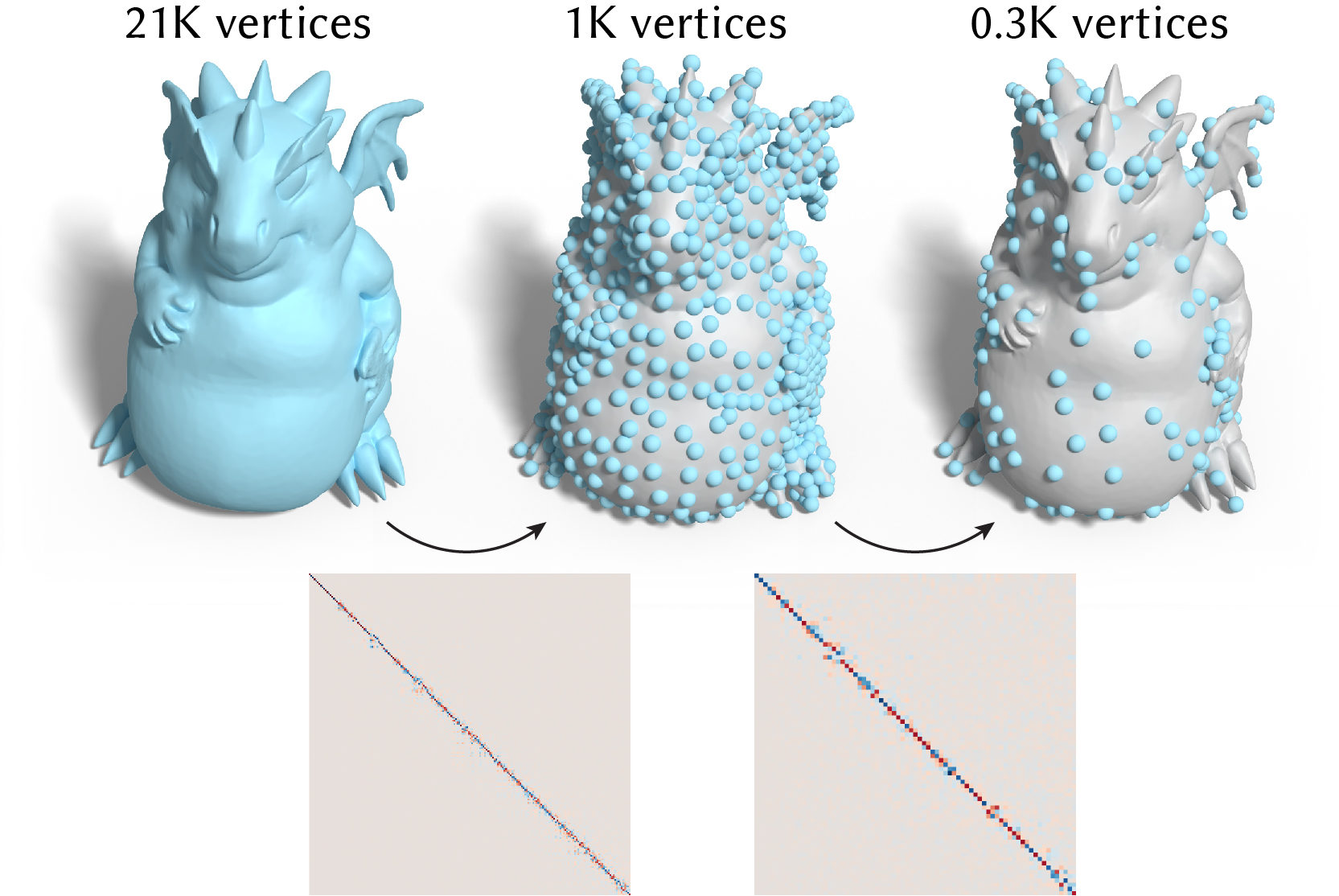}
  \caption{We apply our approach recursively to construct a multilevel hierarchy: from 21,000 rows through 1,000 rows to finally 300 rows.}
  \label{fig:multilevel}
  \vspace{-5pt}
\end{figure}

\update{Our algebraic approach takes the operator as the input, instead of the mesh, thus the output quality is robust to noise or sharp features (see \reffig{meshNoise}). In addition, we can apply our method recursively to the output operator to construct a multilevel hierarchy (see \reffig{multilevel}).} 

\subsection{Comparisons} 
\begin{figure}
  \centering
  \includegraphics[width=3.33in]{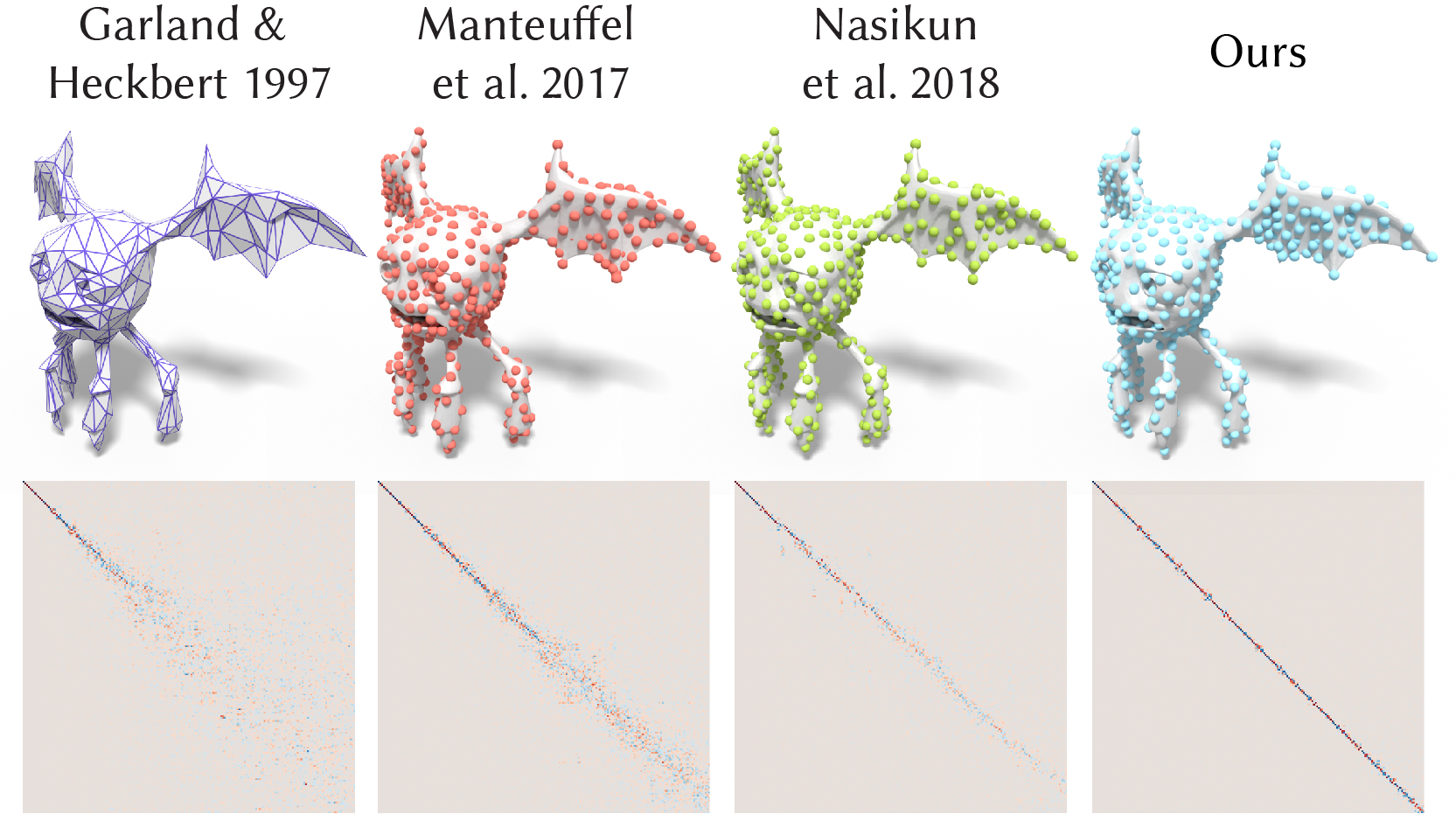}
  \caption{We simplify the cotangent Laplacian from $n=30,000$ to $m=500$.
  Our coarsening preserve the first 200 eigenfunctions better than the QSlim
  \cite{garland1997surface}, the root-node algebraic multigrid
  \cite{manteuffel2017root}, and the fast approximation \cite{nasikun2018fast}.}
  \label{fig:comparison}
  \vspace{-10pt}
\end{figure}

\begin{figure}
  \centering
  \includegraphics[width=3.33in]{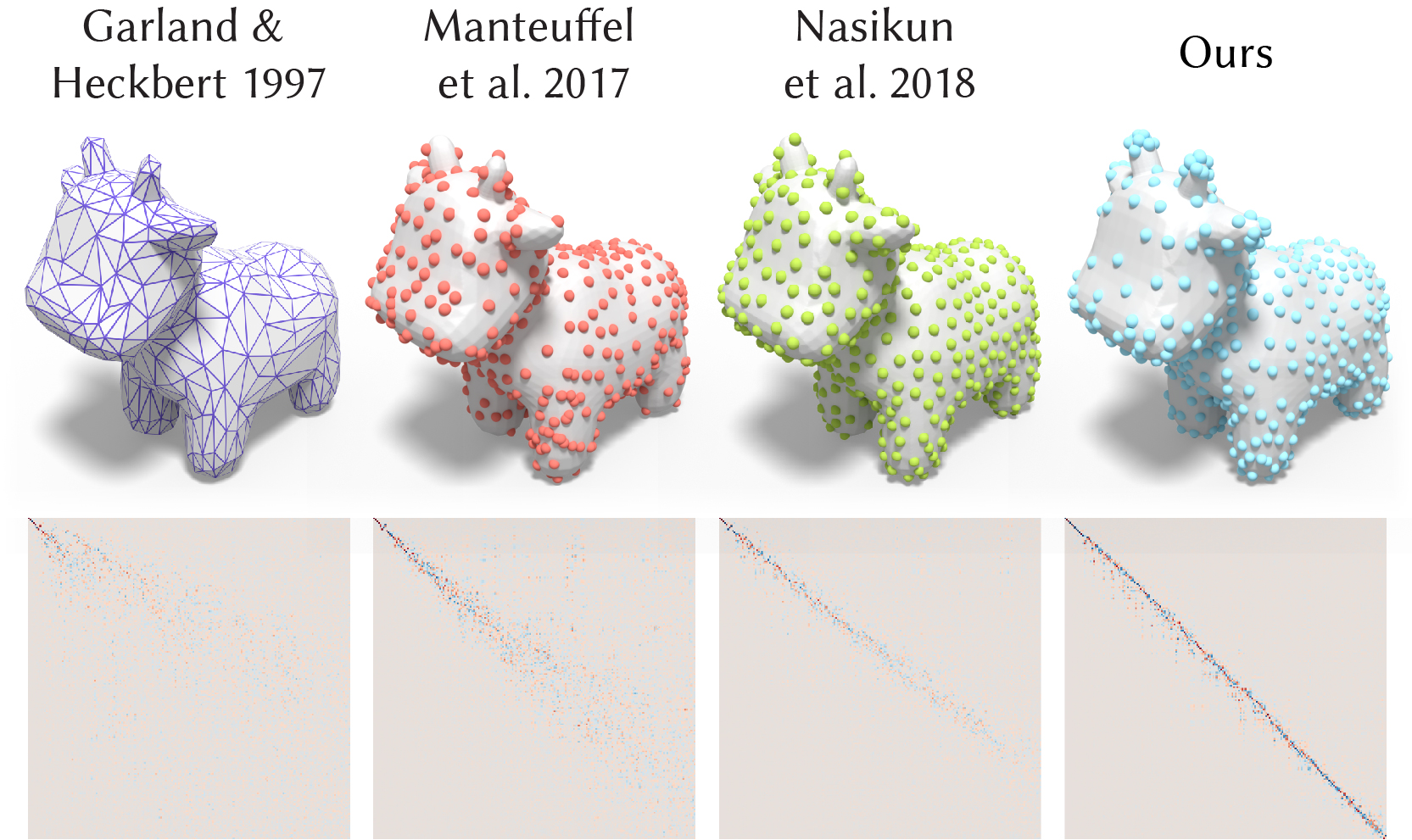}
  \caption{We simplify the anisotropic Laplacian \cite{andreux2014anisotropic}
  (with parameter 70) from $n=25,000$ to $m=600$. Our approach can
  preserve eigenfunctions of anisotropic operators better than the existing
  approaches.}
  \vspace{-5pt}
  \label{fig:comparison_aniso}
\end{figure}

\update{Existing coarsening methods are usually not designed for preserving the spectral property of operators.}
Geometry-based mesh decimation (i.e., \textit{QSlim} \cite{garland1997surface})
is formulated to preserve the appearance of the geometry, and results in poor performance in preserving the operator (see
\reffig{teaser}). As an
iterative solver, algebraic multigrid, i.e., root-node method
\cite{manteuffel2017root}, optimizes the convergence rate \update{and does not preserve the spectral properties} either. Recently, \citet{nasikun2018fast} propose
approximating the isotropic Laplacian based on constructing locally supported
basis functions. However, this approach falls short in \update{preserving the spectral properties of} shapes with
high-curvature thin structures and anisotropic operators (see \reffig{comparison}, \reffig{comparison_aniso}). In contrast, our proposed
method can effectively preserve the eigenfunctions for both isotropic and
anisotropic operators.

\begin{figure}
  \centering
  \includegraphics[width=3.33in]{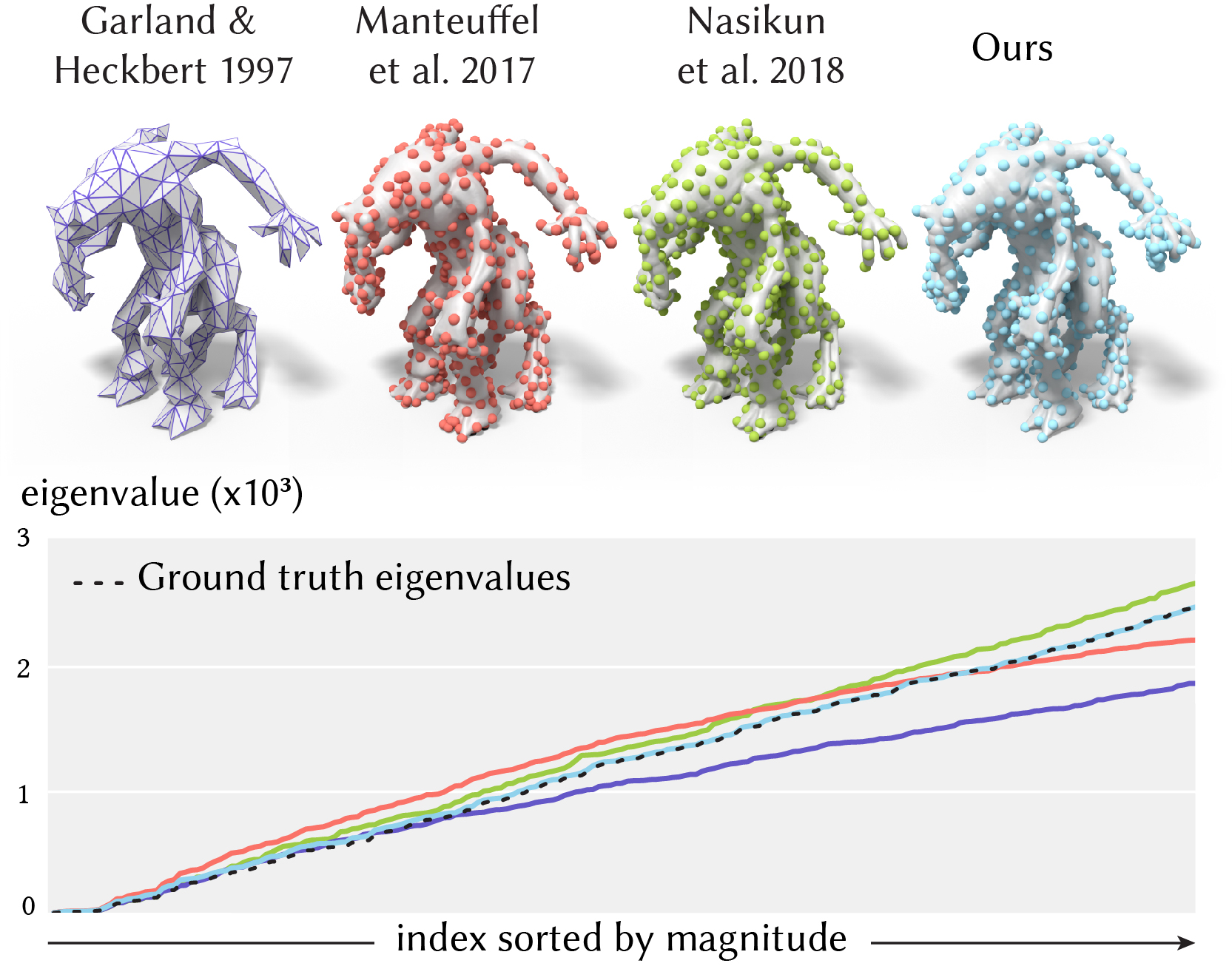}
  \caption{We compare the performance of preserving eigenvalues with different simplification methods. As optimizing our proposed energy implies eigenvalue preservation, we show that the eigenvalues of the simplified operator is well-aligned with the original eigenvalues.
  }
  \label{fig:eVal}
  \vspace{-10pt}
\end{figure}

In addition, a simple derivation \update{(\refapp{eValPreservation})} can show that minimizing the proposed energy implies eigenvalue preservation \update{(see \reffig{teaser_eVal} and \reffig{eVal}).}

\begin{figure}
  \centering
  \includegraphics[width=3.33in]{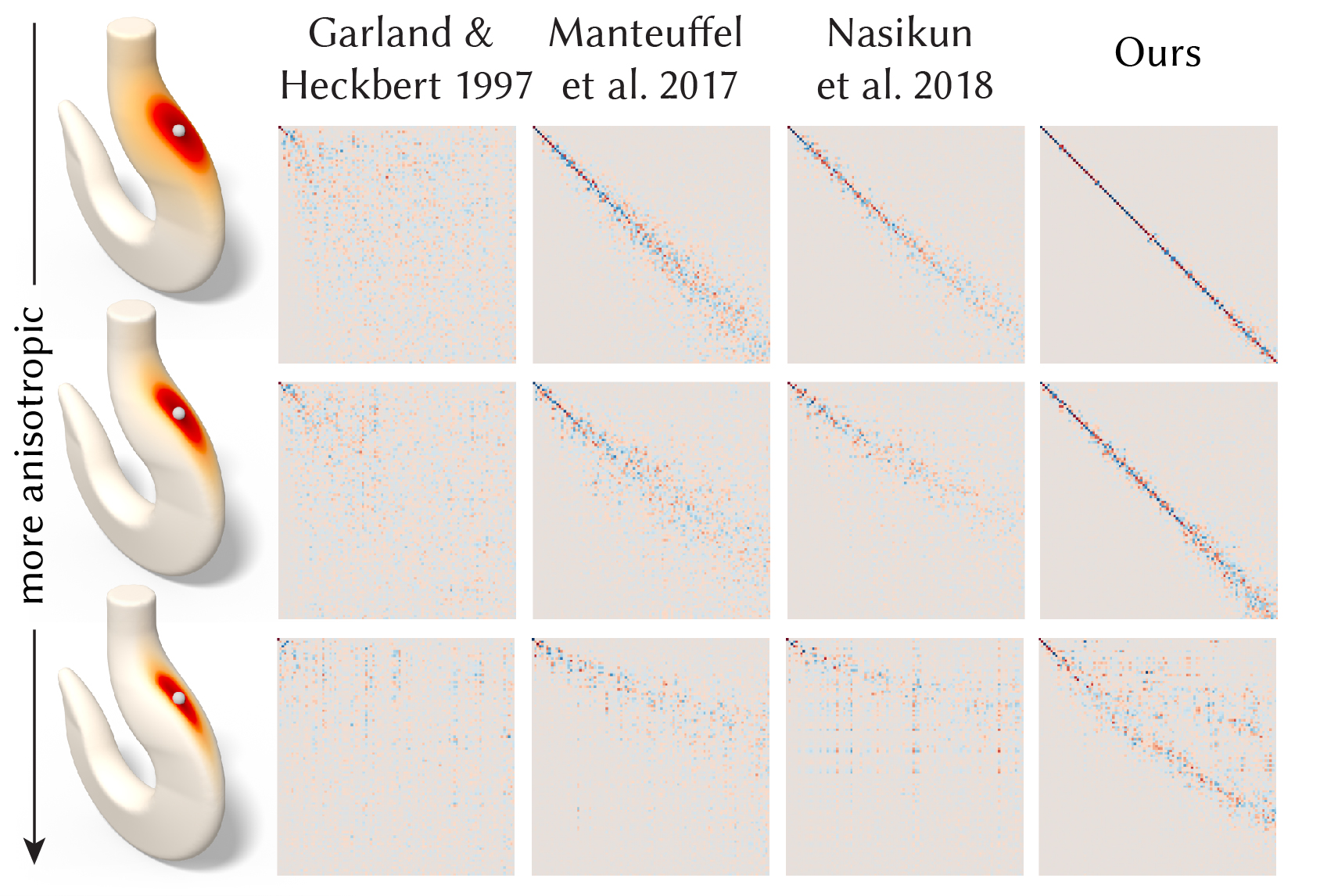}
  \caption{We increase the anisotropy parameter of \cite{andreux2014anisotropic}
  (60, 120, 180)
  while simplifying an operator from 21,000 rows down to 500. Our approach
  handles anisotropy better than existing approaches but still struggles to
  preserve extreme anisotropic operators. }
  \label{fig:extremeAniso}
  \vspace{-5pt}
\end{figure}

In \reffig{extremeAniso}, we show that our method handles anisotropy in the
input operator better than existing methods. This example also demonstrates how
our method \update{gracefully} degrades as anisotropy increases. Extreme anisotropy (far
right column) eventually causes our method to struggle to maintain eigenvectors.

\subsection{Implementation}
\begin{figure}
  \centering
  \includegraphics[width=3.33in]{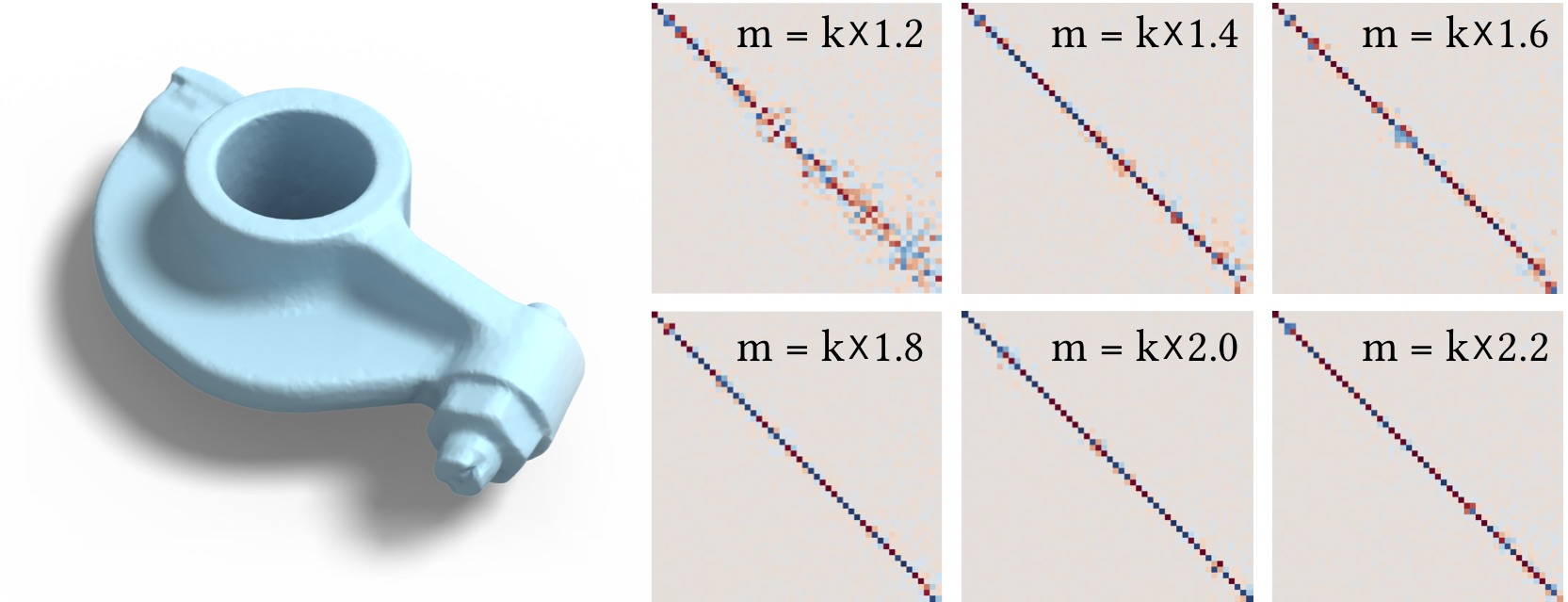}
  \caption{Let $k$ be the number of eigenvectors we want to preserve, experimentally we observed
    that $m>k\times 2$ leads to desired results.}
  \label{fig:spectralSpatial}
  \vspace{-10pt}
\end{figure}

In general, let $k$ be the number of eigenvectors/eigenvalues in use, we recommend to use the number of root nodes $m > k\times 2$. In \reffig{spectralSpatial} we show that if $m$ is too small, the degrees of freedom are insufficient to capture the eigenfunctions with higher frequencies. 

\begin{figure}
  \centering
  \includegraphics[width=3.33in]{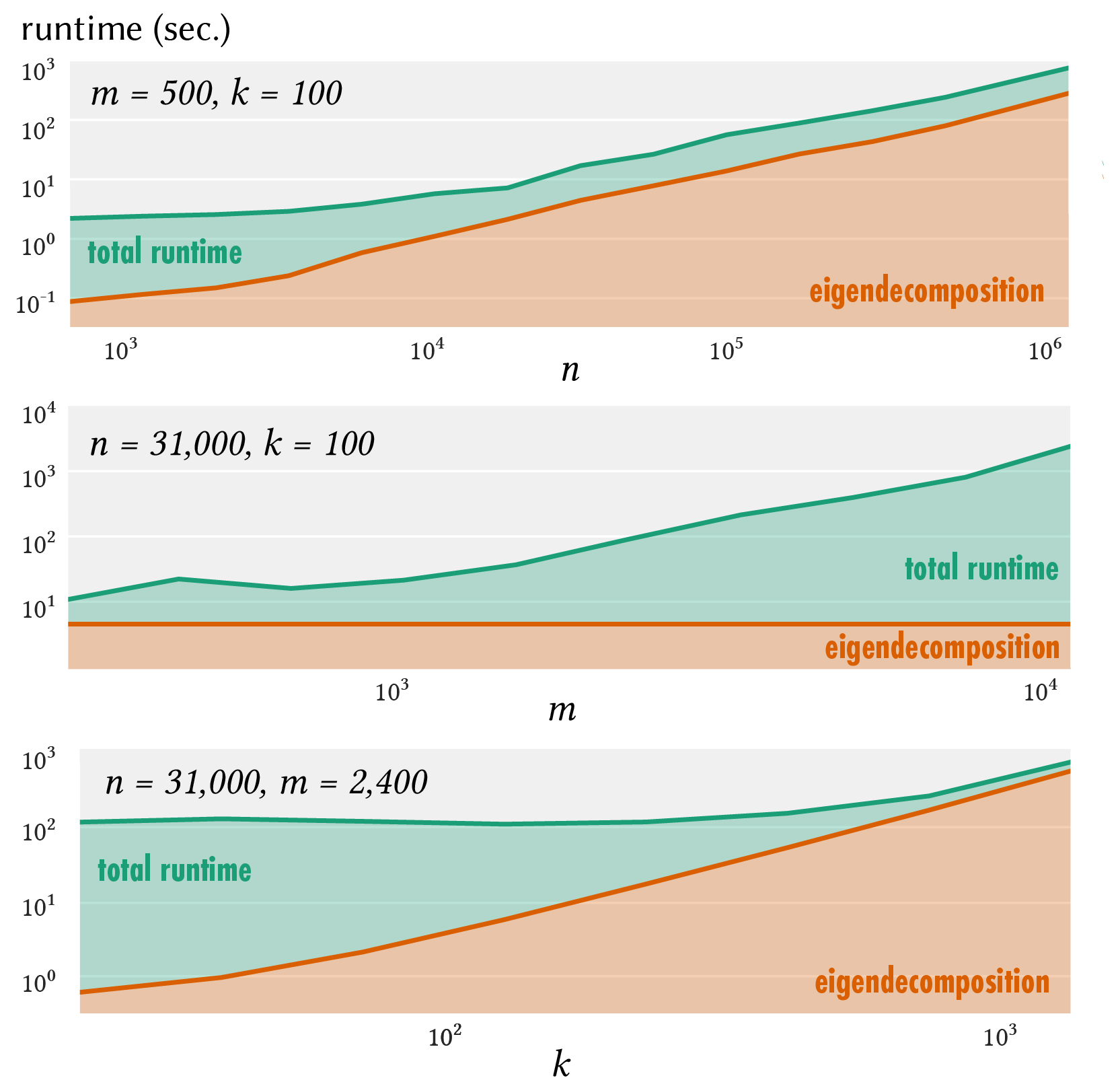}
  \caption{\update{Our runtime shows that our approach is more suitable for aggressive coarsening (middle). For large input meshes and many eigenvectors in use (top, bottom), computing eigendecomposition is the bottleneck.}}
  \label{fig:runtime}
\end{figure}

Our serial C++ implementation is built on top of \textsc{libigl} \cite{libigl} and \textsc{spectra}
\cite{spectra}. \update{We test our implementation on a Linux workstation with an Intel Xeon 3.5GHz CPU,
64GB of RAM, and an NVIDIA GeForce GTX 1080 GPU. We evaluate our runtime using the mesh from \reffig{visualizeFMap} in three different cases: (1) varying the size of input operators $n$, (2) varying the size of output operators $m$, and (3) varying the number of eigenvectors in use $k$.
All experiments converge in 100-300 iterations. We report our runtime in \reffig{runtime}. We obtain 3D shapes mainly from \textit{Thingi10K} \cite{Thingi10K} and clean them with \update{the method of} \cite{hu2018tetrahedral}.}
%


\subsection{Difference-Driven Coarsening}
\begin{figure} 
  \centering
  \includegraphics[width=3.33in]{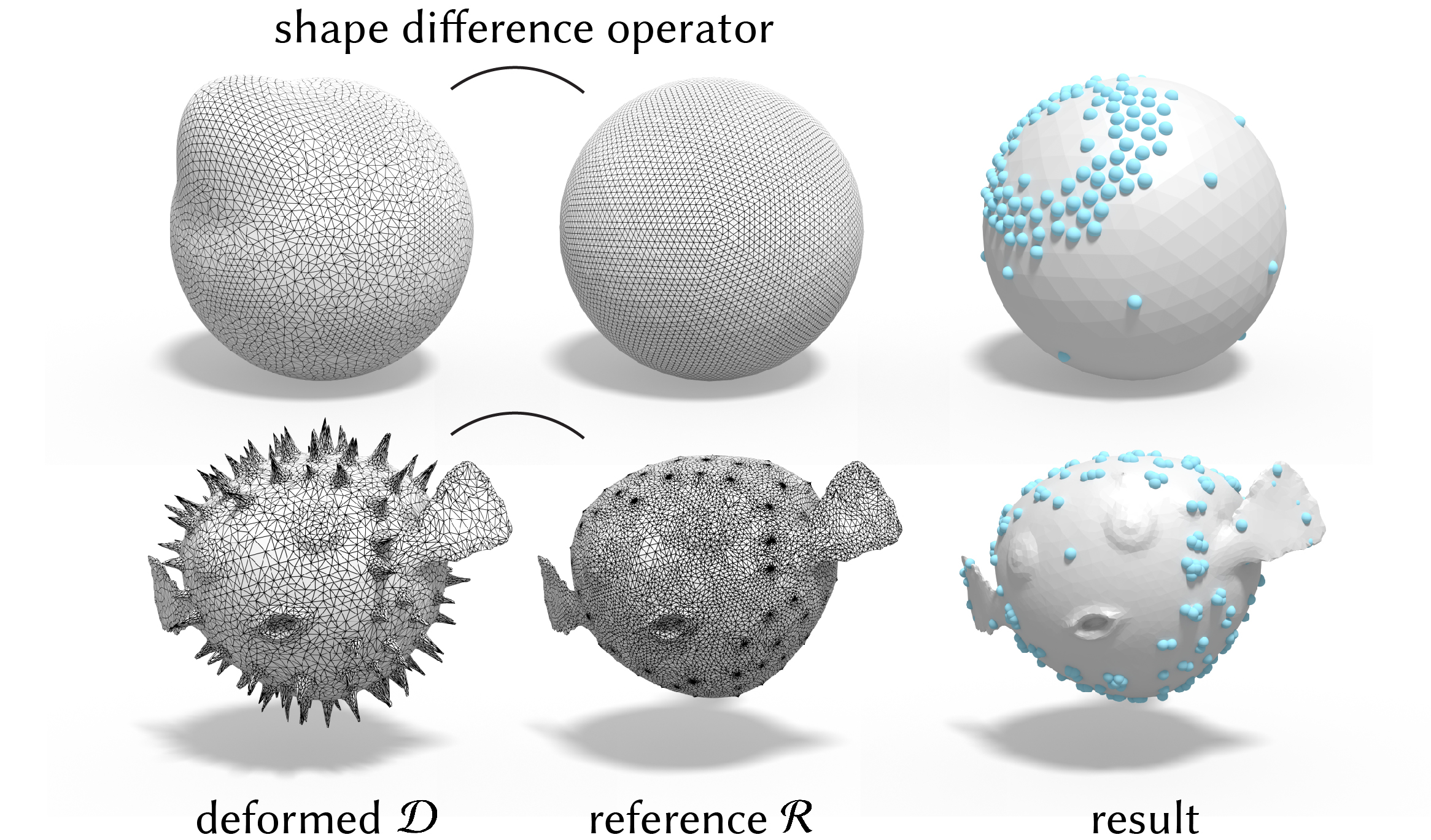}
  \caption{Given a reference shape $\mathcal{R}$ and its deformed version $\mathcal{D}$, we combine the shape difference operator with our coarsening to compute a set of samples that capture the areas of highest distortion between the shapes.}
  \label{fig:SDSimExample} 
  \vspace{-5pt}
\end{figure}

We also validate our combinatorial coarsening by applying it to the shape difference operator \cite{rustamov2013map}
which provides an informative representation of how two shapes differ from each other. As a positive definite operator it fits naturally into our framework. Moreover, since the difference is captured via functional maps, it does not require two shapes to have the same triangulation. We therefore take a pair of shapes with a known functional map between them, compute the shape difference operator and apply our combinatorial coarsening, while trying to best
preserve this computed operator. Intuitively, we expect the samples to be informed by the shape difference and thus
capture the areas of distortion between the shapes (see \refapp{shapeDif} for more detail). As shown in
\reffig{SDSimExample}, our coarsening indeed leads to samples in areas where the intrinsic distortion happens, thus
validating the ability of our approach to capture and reveal the characteristics of the input operator. 

\begin{figure}
  \centering
  \includegraphics[width=3.33in]{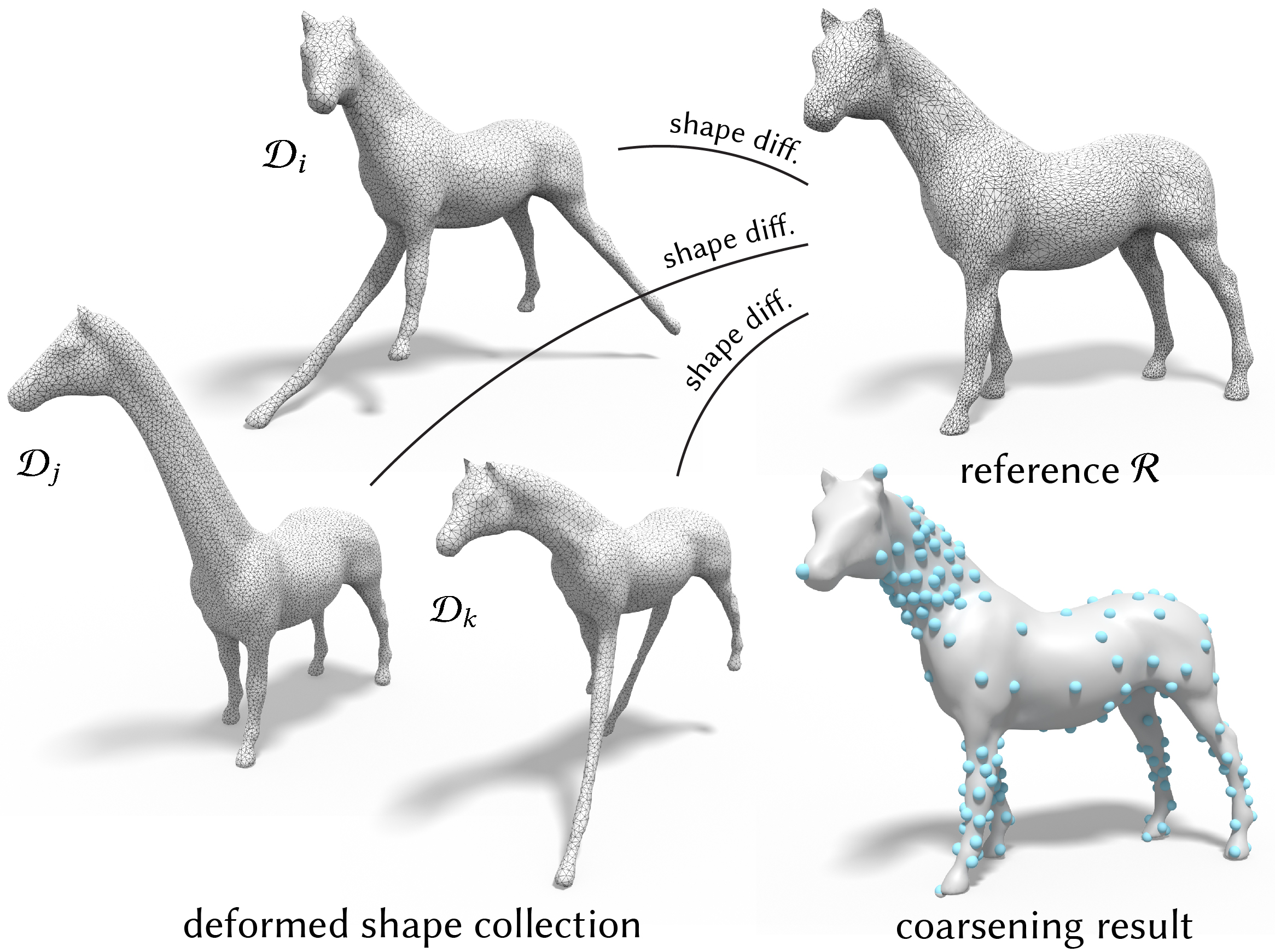}
  \caption{By combining the shape difference operators from the reference shape $\mathcal{R}$ to a collection of deformed shapes $\mathcal{D}$, algebraic coarsening can simplify a mesh based on the ``union'' of all the deformations.}
  \label{fig:SDSim}
\end{figure}

We can further take the element-wise maximum from a collection of shape difference operators to obtain a data-driven coarsening informed by many shape differences (see \reffig{SDSim}).

\subsection{Efficient Shape Correspondence}
\begin{figure}
  \centering
  \includegraphics[width=3.33in]{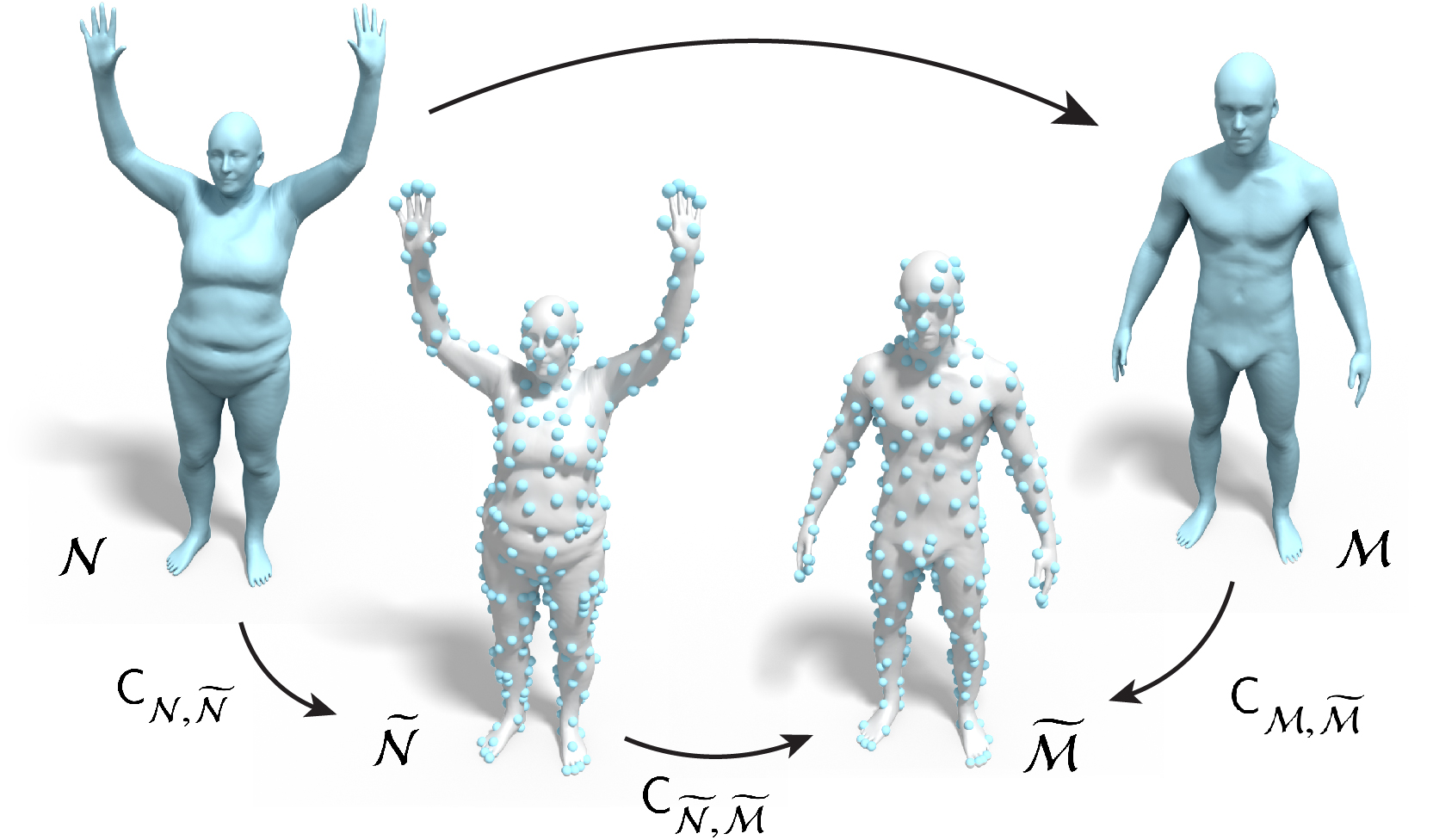}
  \caption{\update{Efficient matching computes} the map ${\C_{\src, \tar}}$ between the original shapes by (1) applying our proposed coarsening to the shape pair and obtain $\C_{\src, \srcc}, \C_{\tar, \tarc}$, (2) compute shape correspondences ${\C_{\srcc, \tarc}}$ in the reduced space, and (3) solve a linear system based on the commutative diagram.}
  \label{fig:hierFMap}
  \vspace{-5pt}
\end{figure}
A key problem in shape analysis is computing correspondences between pairs of non-rigid shapes. In
this application we show how our coarsening can significantly speed up existing shape
matching methods while also leading to comparable or even higher accuracy. For this we use a recent
iterative method based on the notion of \emph{Product Manifold Filter} (PMF), which has shown excellent
results in different shape matching applications \cite{matthias2017kernel}. This method, however,
suffers from high computational complexity, since it is based on solving a linear assignment problem\update{, $O(n^3)$,}
at each iteration. \update{Moreover, it requires} the pair of shapes to have the same number of vertices. As
a result, in practice before applying PMF shapes are typically subsampled to a coarser resolution
and the result is then propagated back to the original meshes. For example in \cite{litany2017deep},
the authors used the standard QSlim \cite{garland1997surface} to simplify the meshes before matching
them using PMF. Unfortunately, since standard appearance-based simplification
methods can severely distort the spectral properties this can cause problems for spectral methods
such as \cite{matthias2017kernel} both during matching between coarse domains and while propagating
back to the dense ones. Instead our spectral-based coarsening, while not resulting in a mesh
provides all the necessary information to apply a \update{spectral} technique via the eigen-pairs of
the coarse operator, and moreover provides an accurate way to propagate the information back to the
original shapes.

More concretely, we aim to find correspondences between the coarsened shapes $\srcc, \tarc$ and to
propagate the result back to the original domains $\src, \tar$ by following a commutative diagram
(see \reffig{hierFMap}). When all correspondences are encoded as functional maps this diagam can be
written in matrix form as:
\begin{align}
  \C_{\tar, \tarc} \C_{\src, \tar} = \C_{\srcc, \tarc} \C_{\src, \srcc},
  \label{equ:HFMapLinearSystem}
\end{align}
where $\C_{\mathcal{X}, \mathcal{Y}}$ denotes the functional map from $\mathcal{X}$ to
$\mathcal{Y}$. Using Eq. \ref{equ:HFMapLinearSystem}, the functional map $\C_{\src, \tar}$ can be
computed by solving a simple least squares problem, via a single linear solve. Our main observation
is that if the original function space is preserved during the coarsening, less error will be
introduced when moving across domains.
\begin{figure} 
  \centering
  \includegraphics[width=3.33in]{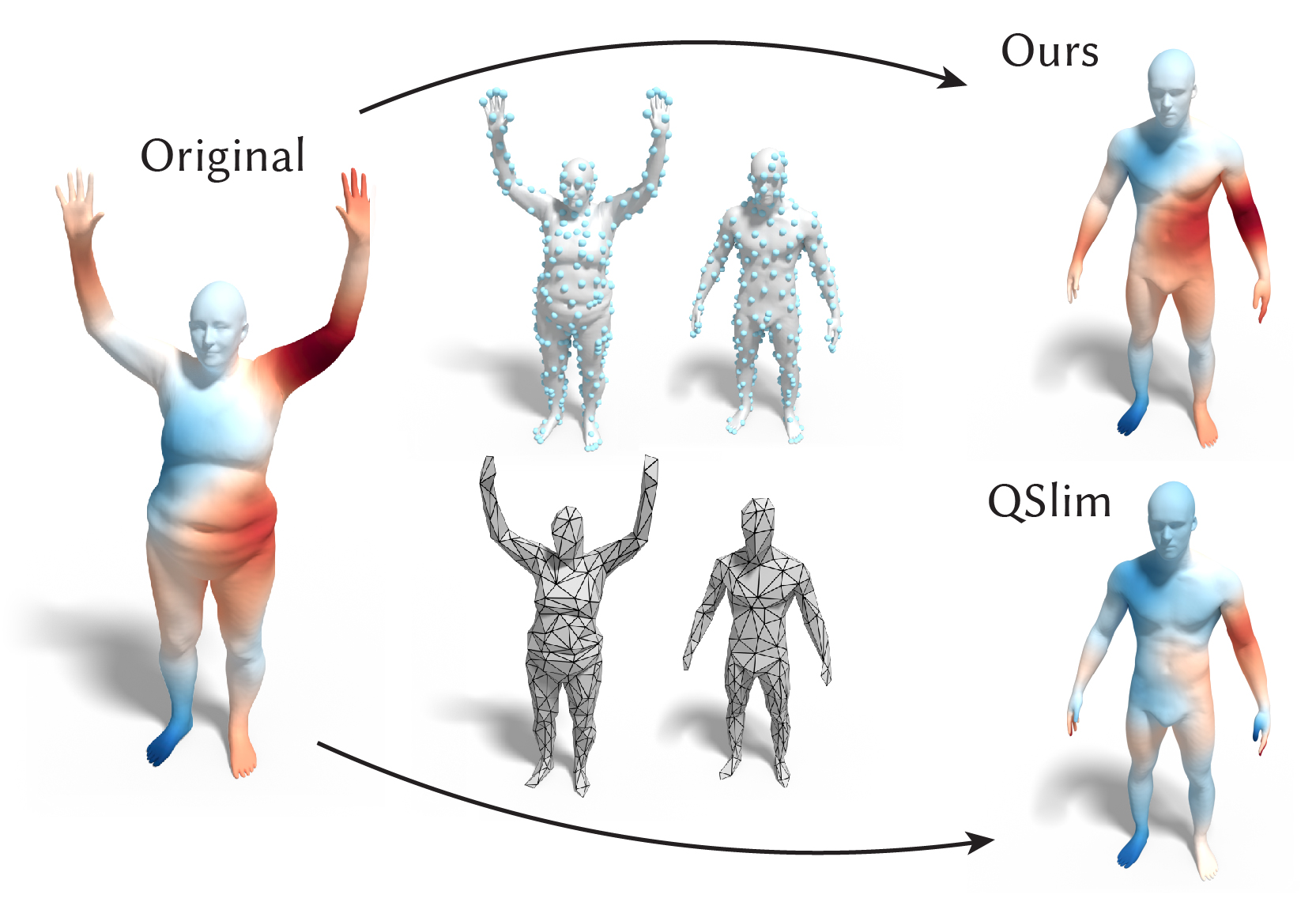}
  \caption{\update{Using our coarsening (top) to infer functional maps between the original pair from the coarse pair introduces less error than using the appearance-based mesh simplification (bottom), QSlim \cite{garland1997surface}.}}
  \label{fig:hierFMap_QSlim}
  \vspace{-10pt}
\end{figure}

\begin{figure}
  \centering
  \includegraphics[width=3.33in]{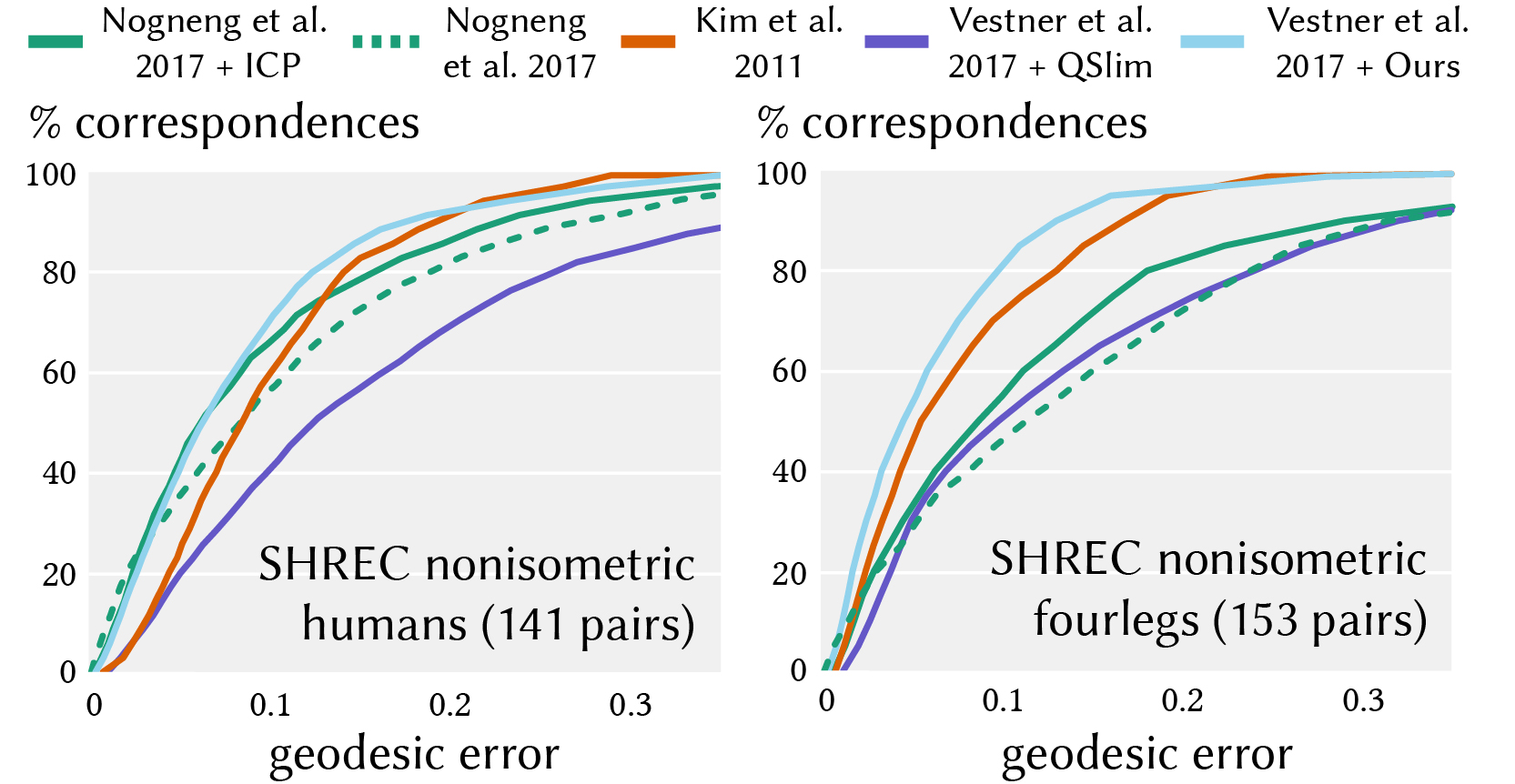}
  \caption{Our \update{efficient shape correspondence} with kernel matching \cite{matthias2017kernel} achieves comparable matching quality on many non-isometric shape pairs from \textsc{SHREC} \cite{Giorgi_shaperetrieval} dataset with methods that directly operator on dense meshes \cite{kim2011blended, nogneng2017informative}.}
  \label{fig:hierFMapQuan}
  \vspace{-10pt}
\end{figure}

We tested this approach by evaluating a combination of our coarsening with \cite{matthias2017kernel}
and compared it to several baselines on a challenging non-rigid non-isometric dataset containing
shapes from the SHREC 2007 contest \cite{Giorgi_shaperetrieval}, and evaluated the results using the
landmarks and evaluation protocol from \cite{kim2011blended} (please see the details on both the
exact parameters and the evaluation in the Appendix). Figure \ref{fig:hierFMap_QSlim} shows
the accuracy of several methods, both that directly operate on the dense meshes
\cite{kim2011blended, nogneng2017informative} as well as using kernel matching
\cite{matthias2017kernel} with QSlim and with our coarsening. The results in \update{Figure
\ref{fig:hierFMap_QSlim} show} that our approach produces maps with comparable quality or superior
quality to existing methods on these non-isometric shape pairs, and results in significant
improvement compared to coarsening the shapes with QSlim. At the same time, in Table
\ref{tab:runtime} we report the runtime of different methods, which shows that our approach \update{leads to} a
significant speed-up compared to existing techniques, and enables an efficient and accurate
PMF-based matching method
(see
\reffig{hierFMapQuan}) with significantly speedup.
\begin{table}[b] 
  \centering
  \caption{
    We report the total time, pre-computation time + runtime, for computing a 60-by-60 functional map on a shape pair with 14,000 vertices each. Our pre-computation time will be amortized by the number of pairs because we apply coarsening on each shape independently, and the number of combinations is quadratic in the number of shapes. Our runtime is orders of magnitude faster because we only need to perform shape matching in the coarsened domain (i.e., 300 vertices).
  }
  \vspace{-5pt}
  \begin{tabularx}{\linewidth}{cccc}
  \toprule
    \textit{[Nogneng 17]+ICP} & \textit{[Nogneng 17]} & \textit{[Kim 11]} & \textit{[Vestner 17]+ours} \\
    32.4 $sec$ & 4.6 $sec$ & 90.6 $sec$ & 10.8+\textbf{0.3} $sec$\\
  \bottomrule
  \end{tabularx}
  \smallskip
  \label{tab:runtime}
\end{table}

\subsection{Graph Pooling} \label{sec:graphPooling}
Convolutional neural networks \cite{lecun1998gradient} \update{have} led to breakthroughs in image, video, and sound recognition tasks. The success of CNNs has inspired a growth of interest in generalizing CNNs to graphs and curved surfaces \cite{bronstein2017geometric}. The fundamental components of a graph CNN are the \emph{pooling} and the \emph{convolution}. Our root node representation $\P, \K$ defines a way of performing pooling on graphs. Meanwhile, our output $\Lc$ facilitates graph convolution on the coarsened graph due to the convolution theorem \cite{arfken1999mathematical}.

\begin{figure}
  \centering
  \includegraphics[width=3.33in]{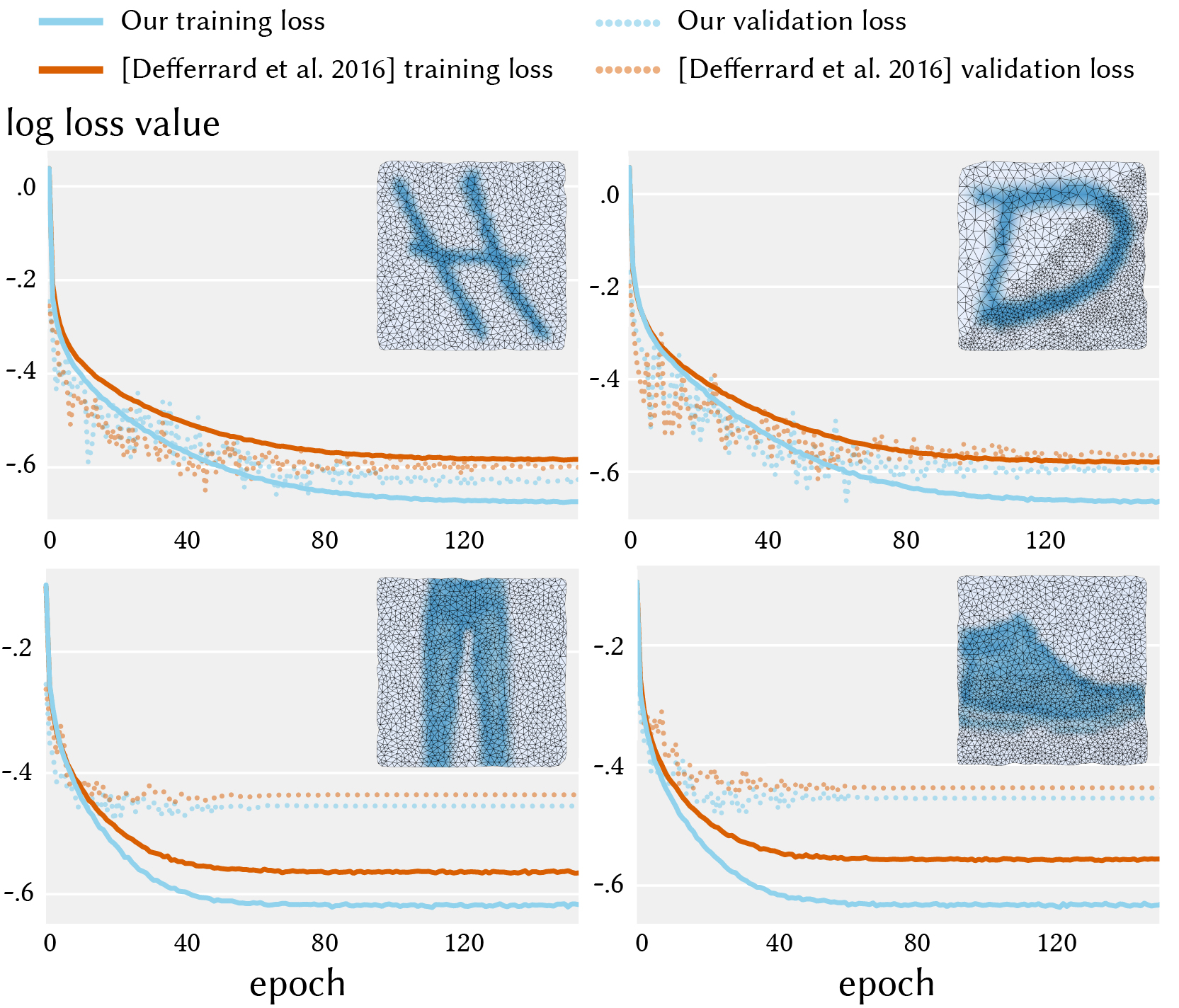}
  \caption{Graph pooling using our coarsening performs better than the pooling presented in \cite{defferrard2016convolutional} on classifying the mesh EMNIST (top row) and the mesh fashion-MNIST (bottom row) datasets.}
  \label{fig:graphPooling}
  \vspace{-10pt}
\end{figure}

To evaluate the performance of graph pooling, we construct several mesh EMNIST datasets where each mesh EMNIST digit is stored as a real-value function on a triangle mesh. Each mesh EMNIST dataset is constructed by overlaying a triangle mesh with the original EMNIST letters \cite{cohen2017emnist}. We compare our graph pooling with the graph pooling IN \cite{defferrard2016convolutional} by \update{evaluating} the classification performance. For the sake of fair comparisons, \update{we use the same graph Laplacian, the same architecture, and the smae hyperparameters.} The only difference is the graph pooling module. In addition to EMNIST, we evaluate the performance on the fashion-MNIST dataset \cite{xiao2017online} under the same \update{settings}. In \reffig{graphPooling}, we show that our graph pooling results in better training and testing performance. We provide implementation \update{details} in \refapp{graphPoolingDetail}. 
\vspace{-5pt}
\section{Limitations \& Future Work}
Reconstructing a valid mesh from our output coarsened operator would enable more downstream applications.
\update{Incorporating a fast eigen-approximation or removing the use of eigen decomposition would further scale the spectral coarsening.
Moreover, exploring sparse SDP methods (e.g. \cite{sun_2015}) could improve our operator optimization.
Jointly optimizing the sparsity and the operator entries may lead to even better solutions.
}
Further restricting the sparsity pattern of the coarsened
operator while maintaining the performance would aid to the construction of a deeper multilevel representation, \update{which could aid in developing a hierarchical graph representation for graph neural networks.}
A scalable coarsening with a deeper multilevel
representation could promote a multigrid solver for geometry processing
applications. 
Generalizing to quaternions would extend the realm of our
coarsening to the fields that deal with quaternionic operators. 

\update{ 
\section{Acknowledgments}
This work is funded in part by by a Google Focused Research Award, KAUST OSR Award (OSR-CRG2017-3426), ERC Starting Grant No. 758800 (EXPROTEA), NSERC Discovery Grants (RGPIN2017-05235), NSERC DAS (RGPAS-2017-507938), the Canada Research Chairs Program, Fields Institute CQAM Labs, Mitacs Globalink, and gifts from Autodesk, Adobe, MESH, and NVIDIA Corporation. We thank members of Dynamic Graphics Project at the University of Toronto and STREAM group at the \'Ecole Polytechnique for discussions and draft reviews; Eitan Grinspun, Etienne Corman, Jean-Michel Roufosse, Maxime Kirgo, and Silvia Sell\'an for proofreading; Ahmad Nasikun, Desai Chen, Dingzeyu Li, Gavin Barill, Jing Ren, and Ruqi Huang for sharing implementations and results; Zih-Yin Chen for early discussions. 
}

\bibliographystyle{ACM-Reference-Format}
\bibliography{OptPreserve}
\appendix

\section{Derivative with Respect to Sparse $\G$}\label{app:dEdG}
\update{To use a gradient-based solver in \refalg{PGD} to solve the optimization problem in \refequ{quartic} we need derivatives with respect to the non-zeros in $\G$
(the sparsity of $\SG$). We start with the dense gradient:
\begin{align*}
	\frac{\partial E_k}{\partial \G} &= \frac{\partial }{\partial \G} \frac{1}{2}\|\P
	\M^{-1} \L \Phi - \Mc^{-1} \G^\top \L \G \P \Phi
	\|^2_{\Mc}.
\end{align*}
We start the derivation by introducing two constant variables $\mat{A}, \mat{B}$ to simplify the expression 
\begin{align*}
	&\frac{\partial E_k}{\partial \G} = \frac{\partial }{\partial \G} \frac{1}{2} \|\mat{A}- \Mc^{-1} \G^\top \L \G \mat{B}	\|^2_{\Mc} \\
	&\mat{A} = \P \M^{-1} \L \Phi,\ \mat{B} = \P \Phi.
\end{align*}
Using the fact that $\lap,\mass,\massc$ are symmetric matrices and the rules in matrix trace derivative, 
we expand the equation as follows.
\begin{align*}
	\frac{\partial E_k}{\partial \G} &= \frac{\partial }{\partial \G} \frac{1}{2} \|\mat{A}- \Mc^{-1} \G^\top \L \G \mat{B}	\|^2_{\Mc} \\
	& = \frac{\partial }{\partial \G} \frac{1}{2} \mathop{\text{tr}} \big((\mat{A}^\top - \mat{B}^\top \G^\top \L \G \Mc^{-1} ) \Mc (\mat{A}- \Mc^{-1} \G^\top \L \G \mat{B})\big) \\
	&= -\frac{\partial }{\partial \G} \Big(\mathop{\text{tr}}\big( \mat{B}^\top \G^\top \L \G \mat{A} \big) + \frac{1}{2} \mathop{\text{tr}}\big(  \mat{B}^\top \G^\top \L \G \Mc^{-1} \G^\top \L \G \mat{B}\big)\Big) \\
	&= -\ (\L \G \mat{A} \mat{B}^\top + \L \G \mat{B} \mat{A}^\top)  \\
	&\qquad + (\L \G \Mc^{-1} \G^\top \L \G\mat{B} \mat{B}^\top + \L \G \mat{B} \mat{B}^\top \G^\top \L \G\Mc^{-1}) \\
	&=  \L \G ( -\mat{A} \mat{B}^\top - \mat{B} \mat{A}^\top + \Mc^{-1} \G^\top \L \G\mat{B} \mat{B}^\top + \mat{B} \mat{B}^\top \G^\top \L \G\Mc^{-1})
\end{align*}
%
Computing the $\nicefrac{\partial E_k}{\partial \G}$ subject to the sparsity $\SG$ can be naively achieved by first computing the dense gradient $\nicefrac{\partial E_k}{\partial \G}$ and then project to the sparsity constraint through an element-wise product with the sparsity $\SG$. However, the naive computation would waste a large amount of computational resources on computing gradient values that do not satisfy the sparsity. We incorporate the sparsity and compute gradients only for the non-zero entries as
\begin{align*}
	&\Big( \frac{\partial E_k}{\partial \G}\Big)_{ij} = Y_{i*}Z_{*j},\quad i,j \in \SG\\
	&Y = \L\G\\
	&Z = -\mat{A} \mat{B}^\top - \mat{B} \mat{A}^\top + \Mc^{-1} \G^\top \L \G\mat{B} \mat{B}^\top + \mat{B} \mat{B}^\top \G^\top \L \G\Mc^{-1},
\end{align*}
where $Y_{i*}$, $Z_{*j}$ denote the $i$th row of $Y$ and the $j$th column of $Z$.}

\section{Sparse Orthogonal Projection} \label{app:sparseProjection}
\update{
Let $\vg \in \mathbb{R}^z$ be the vector of non-zeros in $\G$, so that
$\vectorize(\G) = \Z \vg$, where $\Z \in \{0,1\}^{mn×z}$ scatter matrix.

Given some $\G_1$ that does not satisfy our constraints, we would like to find its closest projection onto the matrices that do satisfy the constraints. In other words, we aim at solving:
\begin{align*}
	&\mathop{\text{minimize}}_{\G ≐ \SG} \|\G - \G_1\| \\
	&\text{ subject to } \G \P {\Phi}_0- {\Phi}_0 = 0.
\end{align*}
%
%
Using properties of the vectorization and the Kronecker product, we can now write this in terms of vectorization:
\begin{align*}
	&\mathop{\text{minimize}}_\vg \|\vg-\vg_1\| \\
	&\text{ subject to }
	((\P {\Phi}_0)^⊤ ⊗ \id_m) \Z \vg
	-
	\vectorize({\Phi}_0) = 0.
\end{align*}
%
%
whose solution is given as:
\begin{align*}
	\vg &= \vg_1 - \A^⊤ (\A\A^⊤)^{-1} \vb \\
	\A &= ((\P {\Phi}_0)^⊤  ⊗ \id_m) \Z\\
	\vb &= ((\P {\Phi}_0)^⊤  ⊗ \id_m) \Z \vg_1
	-
	\vectorize({\Phi}_0).
\end{align*}
This can be simplified to an element-wise division when ${\Phi}_0$ is a single vector.
}

\section{Eigenvalue Preservation} \label{app:eValPreservation}
Minimizing the commutative energy \refequ{sdp} implies 
\begin{align}
	\label{equ:sdpEqual}
	\P \M^{-1} \L \Phi_i = \Mc^{-1} \Lc \P \Phi_i.
\end{align}
As $\Phi_i$ comes from solving the generalized eigenvalue problem, for every $\Phi_i$ we must have: \update{$\L \Phi_i = \lambda_i
\M \Phi_i.$} Therefore, \refequ{sdpEqual} implies $\lambda_i \P \M^{-1} \Phi_i = \Mc^{-1} \Lc \P \Phi_i,$ which means that $ \P
\Phi_i$ must be an eigenvector of $\Mc^{-1} \Lc $ corresponding to \emph{the same eigenvalue} $\lambda_i$.

\section{Modified Shape Difference} \label{app:shapeDif}
\citet{rustamov2013map} capture geometric distortions by tracking the inner products of real-valued functions induced by transporting these functions from one shape to another one via a functional map \cite{ovsjanikov2012functional}. This formulation allows us to compare shapes with different triangulations and \update{encode} the shape difference between two shapes using a single matrix. Given a functional map $\C$ between a reference shape $\refshape$ and a deformed shape $\defshape$, the area-based shape difference operator $\shapediff$ can be written as (\cite{rustamov2013map} option 2)
\begin{align*}
  \shapediff_{\refshape, \defshape} = \C^\top \C,
\end{align*}
where $\fmap$ is the functional map from functions on $\refshape$ to functions on $\defshape$. The operator $\shapediff_{\refshape, \defshape}$ encodes the difference between $\refshape$ and $\defshape$. Its eigenvectors corresponding to eigenvalues larger than one encode area-expanding regions; its eigenvectors corresponding to eigenvalues smaller than one encode area-shrinking regions.

Motivated by the goal of producing denser samples in the parts undergoing shape changes, no matter the area is expanding or shrinking, our modified shape difference operator $\modshapediff$ has the form  
\begin{align*}
  \modshapediff_{\refshape, \defshape} = (\id - \shapediff_{\refshape, \defshape})^\top (\id - \shapediff_{\refshape, \defshape}).
\end{align*}
This formulation treats area expanding and shrinking equally, the eigenvectors of eigenvalues larger than zero capture where the shapes differ. 

Note that this $\modshapediff$ is a size $k$-by-$k$ matrix where $k$ is the number of bases in use. We map $\modshapediff$ back to the original domain by 
\begin{align*}
  \modshapediff^\text{orig}_{\refshape, \defshape} = \eVec_\refshape \modshapediff_{\refshape, \defshape} \eVec_\refshape^\top.
\end{align*}
Although the $\modshapediff^\text{orig}_{\refshape, \defshape}$ is dense and lots of components do not correspond to any edge in the triangle mesh, the non-zero components corresponding to actual edges contain information induced by the operator. Therefore by extracting the inverse of the off-diagonal components of $\modshapediff^\text{orig}_{\refshape, \defshape}$ that correspond to actual edges as the $-\L_{ij}$, we can obtain a coarsening result induced by shape differences. 

\section{Efficient Shape Correspondence} \label{app:hierFMapDetail}
We obtain dense functional maps from coarse ones by solving
\begin{align}\label{equ:ptMapToFMap_app}
	\C_{\tar, \tarc} \C_{\src, \tar} = \C_{\srcc, \tarc} \C_{\src, \srcc},
\end{align}
where $\C_{\src, \tar}, \fmap_{\tar, \tarc}, \C_{\srcc, \tarc}$ are functional maps represented in the Laplace basis. $\C_{\srcc, \tarc}$ is the functional map of functions stored in the hat basis. To distinguish $\C_{\srcc, \tarc}$ from others, we use $\T_{\srcc, \tarc}$ to represent the map in the hat basis. \refequ{ptMapToFMap_app} can be re-written as  
\begin{align*}
	\Phi_{\tarc} \C_{\tar, \tarc} \C_{\src, \tar} = \T_{\srcc, \tarc} \Phi_{\srcc} \C_{\src, \srcc},
\end{align*}
where $\Phi$ are eigenvectors of the Laplace-Beltrami operator. Then we can solve the dense map by, for example, the \textsc{MATLAB} backslash.
\begin{align*}
	 \C_{\src, \tar} = (\Phi_{\tarc} \C_{\tar, \tarc}) \setminus (\T_{\srcc, \tarc} \Phi_{\srcc} \C_{\src, \srcc}),
\end{align*}
Due to limited computational power, we often use truncated eigenvectors and functional maps. To avoid the truncation error destroys the map inference, we use rectangular wide functional maps for both $\C_{\tar, \tarc}, \C_{\src, \srcc}$ to obtain a smaller squared $\C_{\src, \tar}$. For instance, the experiments in \reffig{hierFMapQuan} use size 120-by-200 for both $\C_{\tar, \tarc}, \C_{\src, \srcc}$, and we only take the top left 120-by-120 block of $\C_{\src, \tar}$ in use.

To compute $\C_{\srcc, \tarc}$ (or $\T_{\srcc, \tarc}$), we normalize the shape to have
surface area 2,500 for numerical reasons, coarsen the shapes down to 500 vertices, and use the
kernel matching \cite{matthias2017kernel} for finding bijective correspondences. We use
$\alpha = 0.01$ (the same notation as \cite{matthias2017kernel}) to weight the pointwise descriptor,
specifically the wave kernel signature \cite{aubry2011wave}; we use time parameter
$0.01× \text{surface area}$ for the heat kernel pairwise descriptors; we use 7 landmarks for
shape pairs in the SHREC dataset.

\section{Graph Pooling Implementation Detail} \label{app:graphPoolingDetail}
We use the LeNet-5 network architecture, the same as the one used in \cite{defferrard2016convolutional}, to test our graph pooling on the mesh EMNIST \cite{cohen2017emnist} and mesh fashion-MNIST \cite{xiao2017online} datasets. Specifically, the network has 32, 64 feature maps at the two convolutional layers respectively, and a fully connected layer attached after the second convolutional layer with size 512. We use dropout probability 0.5, regularization weight 5e-4, initial learning rate 0.05, learning rate decay 0.95, batch size 100, and train for 150 epochs using SGD optimizer with momentum 0.9. The graph filters have the support of 25, and each average pooling reduces the mesh size to roughly $\nicefrac{1}{8}$ of the size before pooling.

Our mesh is generated by the Poisson disk sampling followed by the Delaunay
triangulation and a planar flipping optimization implemented in \textsc{MeshLab}
\cite{cignoni2008meshlab}. We also perform local midpoint upsampling to
construct meshes with non-uniform discretizations. Then EMNIST letters are
``pasted'' to the triangle mesh using linear interpolation. 


\end{document}